\documentclass[12pt]{iopart}
\usepackage{iopams,graphicx}  
\begin{document}

\title[The First Stars]{
Formation of the First Stars}

\author{Volker Bromm}
\address{Department of Astronomy, University 
of Texas, 2511 Speedway, Austin, TX 78712, USA}

\ead{vbromm@astro.as.utexas.edu}
\begin{abstract}
Understanding the formation of the first stars is one of the frontier
topics in modern astrophysics and cosmology. Their emergence signalled
the end of the cosmic dark ages, a few hundred million years after the
Big Bang, leading to a fundamental transformation of the early Universe
through the production of ionizing photons and the initial enrichment 
with heavy chemical elements. We here review the state of our knowledge,
separating the well understood elements of our emerging picture from
those where more work is required. Primordial star formation is unique
in that its initial conditions can be directly inferred from the
$\Lambda$ Cold Dark Matter ($\Lambda$CDM) model of cosmological
structure formation. Combined with gas cooling that is mediated via
molecular hydrogen, one can robustly identify the regions
of primordial star formation, the so-called minihalos, having total
masses of $\sim 10^6 M_{\odot}$ and collapsing at redshifts $z\simeq 20-30$.
Within this framework, a number of studies have defined a preliminary
standard model, with the main result that the first stars were predominantly
massive. This model has recently been modified to include a ubiquitous
mode of fragmentation in the protostellar disks, such that the
typical outcome of primordial star formation may be the formation
of a binary or small multiple stellar system. We will also discuss
extensions to this standard picture due to the presence of dynamically
significant magnetic fields, of heating from self-annihalating WIMP
dark matter, or cosmic rays. We conclude by discussing possible strategies
to empirically test our theoretical models. Foremost among them are
predictions for the upcoming {\it James Webb Space Telescope} (JWST),
to be launched $\sim 2018$, and for ``Stellar Archaeology'', which
probes the abundance pattern in the oldest, most-metal poor stars in our
cosmic neighborhood, thereby constraining the nucleosynthesis inside the first supernovae.
\end{abstract}

\submitto{\RPP}
\maketitle

\section{Introduction}
Elucidating the formation and properties of the first stars, the
still elusive Population~III (Pop~III), lies at the frontier of
modern astrophysics. Their emergence marks the end of the cosmic
dark ages, transforming the early Universe from its state of initial
simplicity into one of ever increasing complexity (Barkana and Loeb 2001).
Pop~III stars were the sources of the first hydrogen-ionizing photons,
thus initiating the extended process of reionization, and of the first chemical elements
heavier than the hydrogen and helium, together with trace amounts of deuterium and lithium, produced in
the Big Bang
(Karlsson {\it et al} 2013).
Furthermore, the first stars may also have been the sites where magnetic
fields reached dynamically significant levels for the first time in
cosmic history. This process of magnetogenesis may have involved a 
combination of Biermann-battery generation, or of primordial seed 
fields, and subsequent amplification through a turbulent dynamo
(Pudritz 1981, Tan and Blackman 2004, Xu {\it et al} 2008, Schleicher {\it et al} 2010, Sur {\it et al} 2010, Turk {\it et al} 2012).

Primordial star formation thus fundamentally changed the conditions in the
early Universe, during its first billion years of existence. Making predictions
for the Pop~III era, to be tested with upcoming telescopes such as the
{\it James Webb Space Telescope} (JWST), or the planned generation of
20-40 m ground based behemoths, such as the Giant Magellan Telescope (GMT), 
the Thirty-Meter Telescope (TMT), and the European Extremely-Large Telescope
(E-ELT), has the awe-inspiring aim of closing the final gap in our
cosmic worldview. There is widespread anticipation that we are just 
entering a golden age of discovery, rendering this review very timely,
at a crucial junction in the field's history (also see the contributions
in Whalen {\it et al} 2010).

Complementary to the direct search for Pop~III stellar systems are ongoing
and planned meter-wavelength radio experiments that aim to detect the
redshifted 21 cm radiation emitted by neutral hydrogen (H\,{\small I}) at
high redshifts (Furlanetto {\it et al} 2006, Barkana and Loeb 2007). Among
those telescopes are the Low Frequency Array (LOFAR), the Murchison Wide-Field
Array (MWA), the Precision Array to Probe the Epoch of Reionization (PAPER),
and, further ahead, the Square Kilometer Array (SKA). In
mapping the distribution of dense H\,{\small I} clouds, the 21 cm experiments
provide constraints on the formation sites of the first stars. Furthermore, the
ionized (H\,{\small II}) regions around massive Pop~III stars copiously produce
Lyman-$\alpha$ recombination photons. They in turn can modify the strength of
the 21 cm signal, through coupling the H\,{\small I} hyperfine structure levels 
to the temperature of the primordial gas, the so-called 
Wouthuysen-Field effect (see Furlanetto {\it et al} 2006 for details).
Cumulatively, the first stars also contribute to the large-angle
polarization of cosmic microwave background (CMB) photons, expressed
in the optical depth to Thomson scattering as measured by the {\it Wilkinson Microwave Anisotropy Probe} (WMAP)
satellite (Kaplinghat {\it et al} 2003, Komatsu {\it et al} 2011).
Another integrated signal from Pop~III stars may be imprinted in the
cosmic infrared background (CIB), both in its amplitude and spectrum of
fluctuations (reviewed in Kashlinsky 2005).

By virtue of serendipity,
we might be able to detect individual Pop~III stars, as opposed
to their cumulative signature, at the moment of their violent deaths,
either as hyper-energetic supernova (SN) explosions (Mackey {\it et al} 2003),
or as gamma-ray bursts (GRBs). The favoured model for the dominant
population of long-duration bursts is the collapse of a rapidly rotating,
massive star into a black hole (see Bloom 2011 for a pedagogical
introduction). Predominantly massive Pop~III stars may thus be viable
GRB progenitors (Bromm and Loeb 2002, 2006), and such GRBs should be
detectable out to very high redshifts (Ciardi and Loeb 2000, Lamb and
Reichart 2000). Intriguingly, GRBs are now known already out to a spectroscopically
confirmed
$z\simeq 8.2$ (Salvaterra {\it et al} 2009, Tanvir {\it et al} 2009), or even
$z\sim 9.4$, with a less secure, photometry-only measurement (Cucchiara {\it et al} 2011). This
clearly demonstrates the tremendous potential of GRBs as probes of the
high-redshift Universe.

From the viewpoint of theoretical astrophysics, the great appeal of
the first stars is that they provide us with an ideal, simplified
laboratory for the otherwise, in the local Universe, extremely complex
process of star formation. One could succinctly summarize the physics
of Pop~III star formation as follows: gravity, the atomic and molecular physics
of the primordial H/He, together with the particle physics of cold
dark matter (CDM). 
The latter provides the initial conditions for
the problem, with parameters that are now known to exquisite precision
in the wake of WMAP (Komatsu {\it et al} 2011) and {\it Planck}. The complex 
magneto-hydrodynamics (MHD) and impact of radiation fields that are important
in present-day star formation may thus, at least initially, be
neglected. The hope is that this simplicity in the physics can largely
compensate for the current lack of direct observational constraints. Given
this methodological state of affairs, it is important to identify, and 
focus on, those aspects of the formation physics that are most robust,
and are in principle amenable to empirical tests.

This review has the following plan. We begin by describing the large-scale,
cosmological context which is responsible for setting the initial and
boundary conditions for our problem (section~2). In the next two chapters,
we discuss the detailed microphysics of the Pop~III star formation process,
first within what has sometimes been termed `standard model' (section~3),
and then proceeding to extensions of this model, such as the impact
of magnetic fields and heating from self-annihilating dark matter particles
(section~4). Subsequently, we briefly address the significant increase
in complexity expected during second-generation star formation, and how
this may result in the more or less gradual transition to normal
star formation, as observed in our local neighborhood (section~5).
We conclude with a brief survey of promising empirical probes, most
of them related to the death of the first stars (section~6). Our focus 
in this review is on work done roughly during the last decade. Some key
milestones in the fascinating history of the first-star field, which
reaches back to the 1950s, have been recounted elsewhere (Bromm and Larson
2004), and will not be repeated here.

In concluding this introduction, we point out a few select reviews and
monographs that nicely complement our current effort. The cosmological
context is further explored in Rees (2000), Barkana and Loeb (2001), and
Loeb (2010). A comprehensive review of the relevant primordial chemistry is provided in
Galli and Palla (2013). Previous reviews of the first stars include Bromm and
Larson (2004), Glover (2005, 2013), and Bromm {\it et al} (2009). Here, it
is instructive to also consider the lessons from present-day
star formation, summarized, e.g., in Larson (2003), Mac~Low and Klessen 
(2004), Stahler and Palla (2004), McKee and Ostriker (2007), and
Zinnecker and Yorke (2007). The feedback from the first stars on
the early intergalactic medium (IGM) is discussed in Ciardi and Ferrara (2005).
The physics of reionization is surveyed in Meiksin (2009), and the heavy
element enrichment in the wake of the first SNe in Karlsson {\it et al} (2013).
Robertson {\it et al} (2010) and Bromm and Yoshida (2011), as well as the monographs by
Loeb and Furlanetto (2013) and Wiklind {\it et al} (2013), review the
related subject of the first galaxies. Finally, the capability
of the JWST to detect the signature of the first stars is comprehensively
treated in Gardner {\it et al} (2006) and Stiavelli (2009).

\section{Cosmological Context}

The $\Lambda$CDM model of cosmological structure formation, calibrated
to high precision by WMAP (Komatsu {\it et al} 2011), and more recently by the {\it Planck} satellite (Planck
Collaboration 2013), has provided us with
a firm framework for the study of the first stars. Generically, within
variants of the CDM model, where larger structures 
are assembled hierarchically through successive mergers of smaller 
building-blocks, the first stars are expected to form in dark matter (DM)
minihalos of typical mass $\sim 10^{6} M_{\odot}$ at redshifts
$z\sim 20 - 30$. To understand this prediction,
we have to consider two basic ingredients: (i) the evolution of the DM
component, and the related formation history of DM halos; and (ii) the thermal
evolution of the primordial, pure H/He gas that falls into those halos.
Let us discuss these ingredients in turn, beginning with the
DM evolution.

\subsection{Basic Physics: Dark Matter and Primordial Gas}
Because we are interested in the earliest phases of structure formation,
we want to consider small, so-called DM minihalos.
Such a minihalo formed where the primordial density 
field was randomly enhanced over the surrounding matter, and where
gravity eventually amplified this perturbation to the point where
it decoupled from the general expansion of the background Universe (the Hubble flow), turned around, and collapsed.
The outcome of this collapse, entraining the baryonic gas with the 
dynamically dominant DM, is a state of ``virial equilibrium''. 
Approximately, one can characterize the virial state by 
an equality between kinetic and gravitational potential energy:
\begin{equation}
\frac{G M_h}{R_{\rm vir}}\sim v_{\rm vir}^2 \mbox{\ ,}
\end{equation}
where $M_h$ is the total halo mass, and $R_{\rm vir}$, $v_{\rm vir}$
are the virial radius and velocity, respectively.
The theory of gravitational instability (see Rees 2000, Loeb 2010)
shows that the halo density at the moment when the collapse has ended, the
virial density, is:
$\rho_{\rm vir}\simeq 200 \rho_b \mbox{\ ,}$
where $\rho_b\simeq 2.5\times 10^{-30} (1+z)^3$~g cm$^{-3}$ is
the density of the background Universe. Often, the simple top-hat
model is used to analytically understand the collapse and 
virialization process (Tegmark {\it et al} 1997). This model assumes that 
there is a spherical overdensity with uniform density, embedded in
the background Universe.
One finds that the virial radius of a minihalo at $z\sim 20$ is 
$R_{\rm vir}\sim 100$~pc, or in general:
\begin{equation}
R_{\rm vir}\simeq 200 \mbox{\ pc\,}
\left(\frac{M_h}{10^6M_{\odot}}\right)^{1/3}\left(\frac{1+z}{10}\right)^{-1}
\left(\frac{\Delta_c}{200}\right)^{-1/3}
\mbox{\ ,}
\end{equation}
where $\Delta_c=\rho_{\rm vir}/\rho_b$ is the overdensity after virialization is (nearly) complete (Barkana and Loeb 2001).

The gas will heat up as a consequence of the collapse, either via
adiabatic compression or due to shock heating.
One can assign a virial temperature to the gas, which
corresponds to the virial velocity of the DM particles, as follows:
$k_{\rm B} T_{\rm vir} \sim m_{\rm H} v_{\rm vir}^2$.
Combining these expressions, one has:
\begin{equation}
T_{\rm vir}\simeq 2\times 10^3 \mbox{\ K\,}
\left(\frac{M_h}{10^6M_{\odot}}\right)^{2/3}\left(\frac{1+z}{20}\right)
\mbox{\ .}
\end{equation}
Therefore, typical gas temperatures in minihalos
are below the threshold, $\sim 10^{4}$~K, for efficient 
cooling due to atomic hydrogen. This is of great significance.
If the gas were unable to cool, there would be no further collapse,
and consequently no gas fragmentation and star formation. The gas would
simply persist in hydrostatic equilibrium, roughly tracing the density
profile of the DM.
Early on, it was realized
that cooling in the low-temperature
primordial gas had to rely on molecular hydrogen (H$_2$) instead (Saslaw
and Zipoy 1967).

Since the thermodynamic behavior of the primordial gas 
is primarily controlled by H$_{2}$ cooling, it is crucial to
understand the non-equilibrium chemistry of H$_{2}$ formation and destruction
(Haiman {\it et al} 1996, Abel {\it et al} 1997, Galli and Palla 1998,
Glover and Abel 2008).
In the absence of dust grains to facilitate their formation,
molecules have to form in the gas phase. The hydrogen molecule possesses
a high degree of symmetry, because it is made up of two identical
atoms. It therefore does not have a permanent electrical dipole moment,
and rotational transitions cannot occur via rapid electric dipole (allowed)
radiation. Instead, radiative transitions can only occur 
via slow (forbidden) magnetic quadrupole radiation. This is the underlying
reason why it is difficult to form H$_2$ simply by the collision of
two H atoms. They can collide and form a short-lived compound system;
however, they cannot radiate away their excess kinetic energy quickly
enough, so that the compound system decays again, leading to the
separation of the two atoms. In the Galactic interstellar medium (ISM),
dust grains serve as catalysts to overcome this difficulty: the grain
can absorb the excess kinetic energy of the H atoms, which are then
free to move on the surface of the grain to pair up with a partner,
resulting in the formation of a molecule (Hollenbach and Salpeter 1971). This
pathway, however, is not available in the early Universe.

The electric dipole transition rule can be satisfied when
a neutral H atom collides with a charged species.
The most important
formation channel turns out to be the sequence: 
${\rm H} + e^{-} \rightarrow {\rm H}^{-} + \gamma$,
followed by
${\rm H}^{-} + {\rm H} \rightarrow {\rm H}_{2} + e^{-}$.
The free electrons act as catalysts, 
and are present as residue
from the epoch of recombination at $z\simeq 1100$, when the Universe had sufficiently
cooled to enable the formation of neutral hydrogen atoms,
or result from collisional ionization in accretion shocks
during the hierarchical build-up of galaxies.
The formation of hydrogen molecules thus ceases when the free
electrons have recombined. 
An alternative formation channel relies on the intermediary H$^{+}_{2}$
with free protons as catalysts. This formation mode is important at very
high redshifts, $z>100$, where CMB photons are still sufficiently energetic
to destroy ${\rm H}^{-}$, whereas H$^{+}_{2}$ is more tightly bound and can
thus already survive (Tegmark {\it et al} 1997).
The H$^{-}$ channel, however, dominates in most circumstances.
Calculations of H$_{2}$ formation
in collapsing DM halos, idealizing the virialization
of dark matter halos in CDM cosmogonies with the top-hat model,
have found a simple
approximate relationship between the asymptotic H$_{2}$ abundance and
virial temperature in the halo (Tegmark {\it et al} 1997),
$f_{\rm H_{2}}
\propto T_{\rm vir}^{1.5}$.
Thus, the deeper the potential well (the larger the virial temperature),
the higher the resulting molecule abundance, and the stronger the
effect of cooling due to the molecules. Put differently: the primordial
gas has first to become sufficiently hot to be able to cool later on.

\begin{figure}
\includegraphics[width=.9\textwidth]{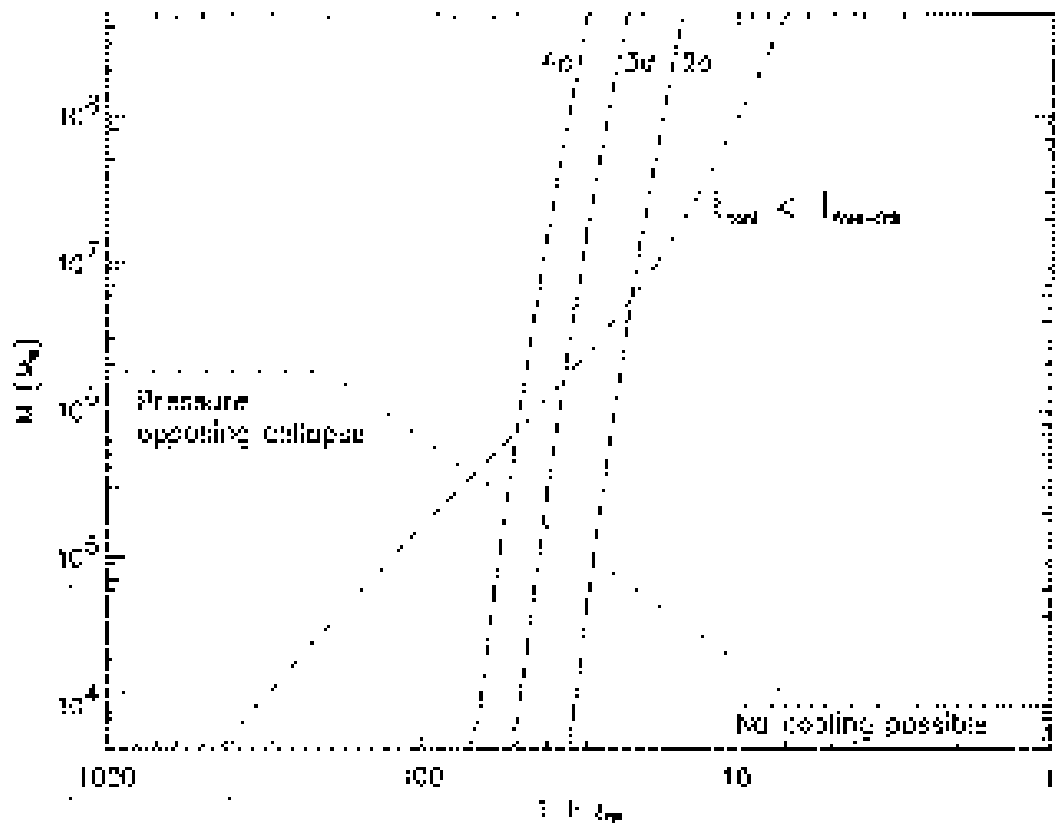}
\caption{Minimum mass of Pop~III star forming region. {\it Solid lines:}
Mass of DM halo vs. collapse redshift for various overdensities.
{\it Dashed line:} Minimum mass to satisfy the Rees--Ostriker--Silk criterion
($t_{\rm cool}<t_{\rm ff}$) vs. redshift. Halos above this line
can cool efficiently and the gas inside them can undergo further collapse.
{\it Dotted line:} Minimum mass to overcome pressure support, effectively
expressing the cosmological Jeans condition (Haiman {\it et al} 1996). Here, it
is assumed that the gas has the same temperature as the CMB for
$z>200$, and evolves adiabatically afterwards. Note that the Jeans criterion 
is not sufficient to enable star formation, and that the stronger cooling
condition has to be met.
For masses below
$\sim 10^4 M_{\odot}$, collapse is not possible since those
halos have virial temperatures below that of the CMB.
Note that the simplified analytical model used here was
adjusted to properly describe the situation at $z\gtrsim 20$, where
cooling and collapse first happen. At later times, its validity breaks down;
in particular it overpredicts the halo masses required for collapse at $z\lesssim 15$
(adopted from Bromm 2000).
}
\label{fig1}
\end{figure}

\subsection{Minimum Halo Mass for Collapse}
Applying the familiar Rees--Ostriker--Silk criterion
for the formation of galaxies
that the cooling timescale has to be shorter than the dynamical
timescale, $t_{\rm cool} < t_{\rm dyn}$, one can derive
the minimum halo mass at a given redshift inside of which the gas
is able to cool and eventually form stars (Rees and Ostriker 1977, Silk 1977).
To carry out this analysis (see figure~1), we need to consider two ingredients:
{\it (i)}: the minimum halo mass that fulfills the Rees--Ostriker--Silk criterion at
a given redshift (dashed line in figure~1); and {\it (ii)}: the $M_h - z_{\rm vir}$ relation,
as given by CDM (solid lines). The latter relation depends on how likely it is that a halo
arises from the gravitational collapse of a peak within
the random field of primordial density fluctuations. Because this field is Gaussian (Loeb 2010), 
one uses the term `$\nu\sigma$'-peak, such that the probability for it to occur scales as
$e^{-\nu/2}$. In the context of determining the formation sites of the first stars, one typically
focuses on $3-4 \sigma$ peaks (Couchman and Rees 1986). Higher $\sigma$ peaks would collapse even
earlier, but they would become too rare to be statistically significant, and would thus not
be able to impact subsequent cosmic history. We further explore this argument in the next
paragraph. The minihalo properties now emerge if one considers the intersection of the two
lines (in figure~1):
a minimum halo mass of $\sim 10^{6} M_{\odot}$
is required for collapse redshifts $z_{\rm vir}\sim 20-30$.
From detailed calculations, one finds that
the gas in such a `successful' halo has reached 
a molecule fraction in excess of
$f_{\rm H_{2}}\sim 10^{-4}$ (Haiman {\it et al} 1996, Tegmark {\it et al} 1997,
Yoshida {\it et al} 2003a).
We note that the H$_{2}$ cooling
function had been quite uncertain pre-1995, differing by an order of magnitude
over the relevant temperature regime.
Advances in the quantum-mechanical
computation of the collisional excitation process (H atoms colliding
with H$_{2}$ molecules) have provided a much more reliable
determination of the H$_{2}$ cooling function (see the discussion and references
in Galli and Palla 1998).

We now consider the connection between halo mass, virialization redshift, and the CDM
power spectrum in somewhat greater detail.
Perturbations on a given mass scale
are characterized by the overdensity
$\delta_M=[(\rho-\rho_b)/\rho_b]_M$,
where the densities are calculated as averages within a spherical
window with a given mass, $M$, inside. Once the overdensity has grown
to a critical value of order unity, $\delta_M=\delta_c\simeq 1.69$,
the perturbation enters its nonlinear phase of collapse and virialization.
Every overdensity can
be expressed as a multiple of the rms fluctuation on the scale in question:
$\delta_M=\nu \sigma(M)$,
which represents a `$\nu \sigma$ peak'.
The rms fluctuation approximately grows
according to:
\begin{equation}
\sigma(M)
\sim \frac{\sigma_0(M)}{1+z} \mbox{\ ,}
\end{equation}
where $\sigma_0(M)$ is the present-day value. For a minihalo with mass
$M_h\sim 10^6 M_{\odot}$, the $\Lambda$CDM model predicts: $\sigma_0(M)
\simeq 15$. One can then estimate the redshift of collapse, or virialization, as:
$1+z_{\rm vir}\sim \nu \sigma_0(M)/\delta_c$,
resulting in $z_{\rm vir}\sim 20 - 30$ for $\nu\sim 3 $.
Reflecting the statistics of
high-$\sigma$ peaks, the primordial clouds
are predominantly clustered, although there are a few cases of
more isolated ones (Yoshida {\it et al} 2003a). As mentioned above,
in principle, DM halos that are sufficiently massive to harbor cold, dense
gas clouds could form at even higher redshifts, $z_{\rm vir}\gtrsim 40$. Such systems, however,
would correspond to extremely rare peaks in the
Gaussian density field (Miralda-Escud\'{e} 2003, Naoz {\it et al} 2006).
Three-dimensional simulations of the combined
evolution of the DM and gas within a cosmological set-up
have confirmed this basic picture (Abel {\it et al} 1998, Bromm {\it et al}
2002, Yoshida {\it et al} 2003a).

As we have seen, whether
a given DM halo successfully hosts a cold, dense
gas cloud
can be nicely understood within the framework of the Rees--Ostriker--Silk criterion (illustrated
in figure~1). Such analytical arguments mainly serve to explain the governing physics in
intuitive terms. A more detailed understanding of minihalo properties has been achieved through a series
of self-consistent cosmological simulations (Fuller and Couchman 2000, Machacek {\it et al} 2001, Yoshida
{\it et al} 2003a, Gao {\it et al} 2007).
To first order, the simulations have confirmed the
analytical derivation, but have added precision and some interesting new effects. Specifically,
the simulations suggest a minimum collapse
mass of $M_{\rm crit}\simeq 7\times 10^{5}M_{\odot}$, with only
a weak dependence on collapse redshift. This somewhat surprising result can be understood
in terms of the merging history of a halo (Yoshida {\it et al} 2003a). 
The dynamical heating accompanying
the merging of DM halos may prevent clouds from successfully cooling 
if they experience too rapid a growth in mass. Halos that virialize later are predicted to
have larger threshold masses, according to the Rees--Ostriker--Silk cooling criterion. This is
partly offset, however, by the reduced dynamical heating experienced in these less dense, and
therefore less biased (clustered) regions
of the cosmological density field.
Subsequently, simulations with even higher numerical resolution have revised the
threshold mass for collapse to somewhat lower values,
$M_{\rm crit}\gtrsim 2\times 10^{5}M_{\odot}$, while confirming the overall picture
(O'Shea and Norman 2007).
The primordial gas clouds that are found in the cosmological
simulations are the sites where the first stars form, further discussed
in the next section.

\begin{figure}
\includegraphics{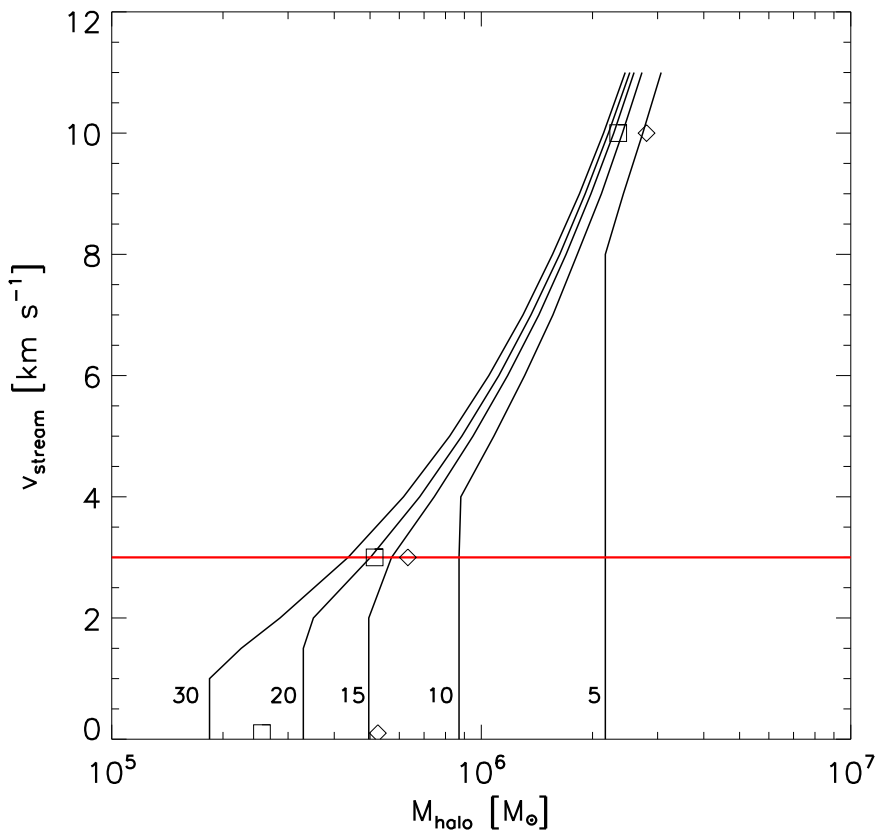}
\caption{Impact of relative baryon-DM streaming on Pop~III star formation.
Shown are relative streaming velocities at $z=100$ vs. minimum halo mass required 
for collapse at different redshifts (as labelled). The solid red line indicates the
rms streaming velocity (Tseliakhovich and Hirata 2010), extrapolated to $z=100$.
The open symbols mark results from select numerical simulations, roughly
in accordance with the analytical results.
The minimum halo
mass increases for larger streaming velocities, and this effect is strongest
at higher redshifts. It is also evident that for small streaming velocities,
there is almost no effect on the halo mass. Note that the vertical portions
for a given $z$ reflect the condition that the decaying streaming motions
are smaller than the sound speed, which only modestly evolves with redshift,
$v_s(z)<c_s$. Once this condition is violated, halo masses have to be larger
to compensate for the added pressure from the surviving streaming motions
(adopted from Stacy {\it et al} 2011a).
}
\label{fig2}
\end{figure}

\subsection{Modifications to the CDM Standard Model}
The initial conditions for Pop~III star formation are set by
the properties of the gravitationally dominant dark matter
component. The `standard model' outlined above is quite robust
to moderate variations of the CDM model parameters. More
profound changes to the underlying particle physics, however,
have noticeable consequences on primordial star formation.
One such modification,
the warm dark matter (WDM) model, leads to a shift in Pop~III formation
sites to more massive halos (Yoshida {\it et al} 2003b), and possibly
into the filamentary structures of the cosmic web (Gao and Theuns 2007).
Recently, interest in WDM scenarios has somewhat waned, however. A second
effect may be of potentially greater significance. This is the
realization that baryons and the CDM fluid engage in a relative,
supersonic streaming motion, coherent on scales of a few comoving Mpc
(Tseliakhovich and Hirata 2010). The baryons before recombination are
tightly coupled to the photon fluid, and are thus forced into an
acoustic oscillation pattern, whereas the CDM fluid does not feel the
presence of the radiation any longer. This results in a relative
baryon-DM velocity of $v_{s,i}\sim 30$~km~s$^{-1}$ at recombination.
At the same time, the baryonic sound speed drops from relativistic
values ($\sim c/\sqrt{3}$) to the thermal velocity of the hydrogen atoms
($\sim 6$~km~s$^{-1}$).

The relative streaming impacts primordial star formation in a number
of ways (Greif {\it et al} 2011b, Maio {\it et al} 2011, Stacy {\it et al}
2011a, Tseliakhovich {\it et al} 2011, Naoz {\it et al} 2012, 2013).
Most importantly, it raises
the minimum halo mass where Pop~III stars can form. This can
be understood as follows. To enable star formation, the halo virial mass
has to be larger than the cosmological Jeans mass:
\begin{equation}
M_{\rm J}\simeq \frac{c_s^3}{G^{3/2}\rho^{1/2}}\mbox{\ ,}
\end{equation}
where $c_s$ is the baryonic sound speed, and $\rho$ the total (DM $+$ gas)
density. To accommodate the presence of the relative streaming, one can
replace the sound speed with an `effective velocity' $v_{\rm eff}=
\sqrt{c_s^2+v_s^2}$, where $v_s=v_s(z)=v_{s,i}[(1+z)/1,000]$ is the redshift-dependent
relative streaming velocity, normalized to the value imprinted at recombination.
The decay with redshift reflects the generic
decay of momentum-energy in an expanding Universe.
As a result, the collapse of a given DM halo is delayed, by $\Delta z \sim$~a few, and at a given redshift, the minimum halo mass for collapse is similarly raised
by a factor of a few. The abundance of Pop~III star-forming minihalos
is thus reduced as well, in turn delaying the onset of reionization. In 
addition, the fluctuations in the H\,{\small I} 21~cm brightness distribution
may be enhanced (Maio {\it et al} 2011, Visbal {\it et al} 2012). We illustrate this
effect in Figure~2. Note that Stacy {\it et al} (2011a) initialize their simulation at $z=100$;
their `initial' streaming velocities are therefore a factor of 10 smaller than the value at
recombination.

\section{Formation Physics: The Standard Model}

\subsection{Basic Processes: Chemistry, Thermodynamics, and Opacity}
During the dissipative collapse in the center 
of a minihalo, the primordial gas experiences three distinct phases,
according to density. For each phase, we will consider the key chemical
and thermal (cooling/heating) processes, as well as the sources of opacity,
if important. Regarding these basic ingredients, we see a substantial
reduction in complexity, as compared to the case of present-day star
formation. However, while we learn more from ever more realistic simulations,
this seeming simplicity is being replaced by an increasingly involved set
of input physics. Indeed, the expectation from a decade ago that the first stars
would provide us with an ideal, simple laboratory for star formation appears somewhat
naive with hindsight. On the other hand, recent progress is an indicator that this
field has entered a much more mature phase, after the initial, pioneering studies.

\subsubsection{Atomic phase}
The evolution below densities of $n\sim 10^8$\,cm$^{-3}$, where the
gas is mostly atomic with only a trace amount of molecular hydrogen
present, is very well understood
(Abel {\it et al} 1997, Galli and Palla 1998). As long as there is
a trace amount of free electrons, either from the residual recombination
abundance or from subsequent ionization events, gas phase reactions will
build up an asymptotic H$_2$ abundance of $n$[H$_2$]$/n\sim 10^{-3}$. This
small molecule fraction suffices to cool the gas, in competition
with the compressional heating during the collapse into the minihalo 
gravitational potential well, to temperatures of about $\sim 200$\,K, reached
at $n\sim 10^4$\,cm$^{-3}$. The latter is the critical density where
the rotational levels of H$_2$ transition from non-LTE to LTE populations.
This transition is sometimes termed `loitering state', as the H$_2$ cooling
rate saturates there, becoming less efficient toward higher densities
(Bromm {\it et al} 2002), or more precisely: changing from a $\Lambda\propto n^2$
dependence to $\propto n$. It has been argued that the characteristic
density and temperature at the loitering state imprint the mass scale of
the Pop~III-equivalent to a pre-stellar core (Bromm {\it et al} 2002). Indeed,
rewriting Equation~(5) for self-gravitating gas, we obtain the Bonnor-Ebert
mass, as follows:
\begin{equation}
M_{\rm J}\simeq 500 M_{\odot}\left(\frac{T}{200{\rm \,K}}\right)^{3/2}
\left(\frac{n}{10^4{\rm \,cm}^{-3}}\right)^{-1/2}\mbox{\ .}
\end{equation}
This pre-stellar core is at the verge of gravitational runaway collapse, and
it is the immediate progenitor of a Pop~III star, or a small multiple thereof,
if further fragmentation happens later on (see section~3.4 below).

The loitering state is reached as long as there is no other low-temperature
coolant besides H$_2$. Such an additional cooling agent could be provided
by hydrogen deuteride (HD), which does possess a permanent electric dipole
moment with a correspondingly larger Einstein-$A$ spontaneous emission
coefficient. Furthermore, HD is able to cool the gas to temperatures
below 200\,K, possibly all the way to that of the CMB, which sets a lower
limit to radiative cooling (Larson 1998, Smith {\it et al} 2009, Schneider and Omukai 2010).
The latter is possible because
of the allowed transition rule $\Delta J=1$, as opposed to
$\Delta J=2$ for the quadrupole transitions in H$_2$, where $J$ is the
rotational quantum number. The HD cooling function, based on state-of-the art quantum-mechanical calculations of the collisional excitation rate, has been provided
in convenient, analytical form (Flower {\it et al} 2000). Including the HD
cooling channel does not change the thermal history of primordial collapse
into a minihalo in most cases (Johnson and Bromm 2006). The only exception
may be the situation in the lowest-mass minihalos, where HD cooling
could facilitate a drop to the CMB temperature floor (McGreer and Bryan 2008).
HD cooling may be of much greater impact in the so-called Pop~III.2 case,
where stars form out of primordial gas that has undergone a significant
degree of pre-ionization (see section~5.1). The higher free-electron abundance, or more
generally the enhanced free-ion fraction, in turn
catalyzes a boost in H$_2$ formation, and indirectly also of HD, as a
consequence of the main formation channel: ${\rm H}_2 + {\rm D}^+ \rightarrow
{\rm HD} + {\rm H}^+$ (Galli and Palla 2002).

The low-density primordial chemistry regime is continuously being refined,
through better quantum-chemistry calculations and laboratory measurements of key rates (Galli and Palla 2013).
A nice example is the high-precision laboratory determination of the
associative detachment (AD) reaction:
${\rm H} + {\rm H}^- \rightarrow
{\rm H}_2 + {\rm e}^-$ (Kreckel {\it et al} 2010). Previously, poor knowledge of the
AD rate had introduced a corresponding uncertainty in the asymptotic
H$_2$ rate, and therefore the temperature of the H$_2$ cooled primordial
gas. Further improvements come from carefully adding hitherto neglected
reactions to the primordial chemistry network. To again give one key
example: Traditionally, primordial chemistry and cooling modules had assumed
that H$_2$ cooling at $n\lesssim 10^8$\,cm$^{-3}$ is facilitated only through excitations from collisions
with neutral H atoms. However, in slightly ionized gas, collisions with
free electrons (and protons) may be important as well (Glover and Abel 2008).

\subsubsection{Molecular phase}
At densities in excess of $\sim 10^8$\,cm$^{-3}$, the primordial gas
is converted into fully molecular form (Palla {\it et al} 1983). In the
absence of dust grains, this is brought about via three-body gas phase
reactions: ${\rm H} + {\rm H} + {\rm H} \rightarrow
{\rm H}_2 + {\rm H}$, followed by
${\rm H} + {\rm H} + {\rm H}_2 \rightarrow
{\rm H}_2 + {\rm H}_2$. The sudden jump in molecule abundance is accompanied
by a corresponding large ($\sim 10^3$) increase in gas cooling. This jump,
however, does not lead to runaway cooling, as there now is a competing
contribution to the heating (but see Greif {\it et al} 2013). The latter arises because molecule formation
releases the sizable binding energy, 4.48\,eV, of H$_2$. The net thermal
effect is a near-isothermal collapse at $T\sim 1,000$\,K. A second effect is 
the softening of the equation of state, from an adiabatic index of
$\gamma_{\rm ad}=5/3$ in the atomic phase to
$\gamma_{\rm ad}=7/5$, reflecting the additional internal degrees of freedom
in the molecules. Theoretical calculations of the 
three-body rates exhibit disconcerting,
order-of-magnitude uncertainties (Turk {\it et al} 2011), leading to correspondingly
huge uncertainties in the high-density thermal evolution. Prospects to remedy
the situation by carrying out laboratory measurements are not good in the foreseeable future (Dan
Savin, priv. comm.). The problem here is that there are no charged particles
involved, compounded with very low rate coefficients, thus rendering the standard charged-beam techniques impotent (see
Kreckel {\it et al} 2010).

At increasingly high densities, $n>10^{12}$\,cm$^{-3}$, the ro-vibrational
lines of H$_2$ become increasingly optically thick. The resulting line
transfer is complex, but recent simulations have managed to forge ahead
with an escape probability method, combined with the Sobolev approximation
(Yoshida {\it et al} 2008, Clark {\it et al} 2011b, Greif {\it et al} 2011a,
2012). Specifically, the following expression for the escape
probability is used:
\begin{equation}
\beta_{\rm esc}=\frac{1-\exp(-\tau)}{\tau}\mbox{\ ,}
\end{equation}
where the line optical depth is $\tau=k_{lu}L_{\rm char}$.
Here, $k_{lu}$ is the absorption coefficient in the given line, and
$L_{\rm char}$ is a length characteristic of the line-formation
region. This is where the Sobolev approximation is invoked, stating that line
photons can escape out of regions that are in- or outflowing, with 
a bulk velocity gradient, $dV_r/dr$, once they have been shifted
out of the central Doppler core: $L_{\rm char}\simeq v_{\rm th}/|dV_r/dr|$.
Here, $v_{\rm th}$ is the thermal (sound) speed, which sets the width of
the Doppler core. The three-dimensional cosmological simulations, using
the probablity escape formalism as described above, can reproduce the
results from one-dimensional fully radiation-hydrodynamics calculations,
giving confidence that the methodology is sound (Omukai and Nishi 1998).

\subsubsection{Approaching protostellar conditions}
Finally, at densities $n> 10^{14}$\,cm$^{-3}$, molecular lines become
optically thick throughout. A final cooling agent now enters the scene,
the H$_{2}$--H$_{2}$ ``supermolecule'', possessing a temporary, van-der-Waals-force induced
electric dipole moment (Frommhold 1994). This gives rise to continuous collision-induced
emission (CIE), and its reverse absorption process (CIA). In most recent
simulations of the Pop~III formation problem, continuum radiative transfer is again
modeled with an escape probability formalism, evaluating the Sobolev
optical depth with an analytical fitting formula for $\tau_{\rm CIE}$ (Ripamonti and Abel 2004).
Alternatively, the continuum optical depth has been calculated with a ray-tracing technique (Hirano and Yoshida 2013),
with the advantage of being able to accommodate realistic three-dimensional geometries, found in disk-dominated 
protostellar accretion flows.
The continuum opacity serves to quickly extinguish the
CIE cooling channel as well. Towards the highest densities ($n>10^{16}$\,cm$^{-3}$), just below the threshold
for forming the intial hydrostatic core at the center of the collapse, the collisional dissociation
of H$_2$ removes the binding energy of 4.48\,eV per molecule from the gas, thereby cooling it. The rise in
central temperature, driven by compressional heating, is thus still limited initially.

Another complication at the highest densities concerns the non-equilibrium chemistry, where a
fully-consistent solution of the coupled network of rate equations becomes prohibitively expensive
in terms of computational cost. The reason is that timesteps drop precipitously, reflecting 
the increasingly short reaction timescales at high $n$: $\Delta t_{\rm sys} < t_{\rm chem}\simeq
n_i/\dot n_i$, where $i$ refers to any of the relevant species. A commonly adopted strategy
is to switch to an equilibrium solver, by setting $\dot n_i=0$ everywhere in the rate equations,
once a pre-set density threshold has been crossed. This is done individually for each
resolution element, a cell in grid-based methods,
or a single particle in smoothed-particle hydrodynamics (SPH). E.g., in Greif {\it et al} (2012), this
switch is carried out at $n_{\rm thresh}=10^{14}$\,cm$^{-3}$.

It will be desirable to simulate the final, partially and completely optically thick,
phases with improved radiative transfer schemes. A promising approach is flux-limited
diffusion (Castor 2004). This technique is still sufficiently inexpensive to be coupled
to fully three-dimensional simulations, as has been done in studies of present-day
star formation (Whitehouse and Bate 2006, Krumholz {\it et al} 2007, 2009).

\subsection{Initial Collapse}

The properties of the initial collapse, until a hydrostatic core first arises
in the center of the DM minihalo, have been robustly 
described, with multiple investigations reaching very similar results
(Abel {\it et al} 2002, Bromm {\it et al} 2002, Bromm and Loeb 2004, Yoshida {\it et al} 2006, 2008,
O'Shea and Norman 2007, Turk {\it et al} 2009, Stacy {\it et al} 2010,
Clark {\it et al} 2011b, Greif {\it et al} 2011a, 2012). At low densities, the gas
is heated through adiabatic compression until reaching $n \sim  1$\,cm$^{-3}$, attaining
maximum temperatures of  $\sim 1,000$\,K, which is of order the virial temperature of the host
minihalo (see equation~3). Afterwards, the gas
begins to cool through H$_2$ rovibrational transitions, reaching
a minimum temperature of $T\sim 200$\,K at a density of
$n\sim  10^4$\,cm$^{-3}$. This is the density at which the gas encounters
the quasi-hydrostatic ‘loitering phase’, already introduced above (see section 3.1.1).
After this phase the gas temperature rises again due to
compressional heating until $n\sim 10^8$\,cm$^{-3}$, when three-body
processes become important and cause the gas to turn
fully molecular by $n\sim 10^{12}$\,cm$^{-3}$. 
Recent, very high-resolution simulations suggest that the chemothermal instability
associated with this transition to fully molecular gas may in turn be able to induce
fragmentation already during the initial collapse, on scales of $\sim 10$\,AU (Greif
{\it et al} 2013). This would be in difference from earlier studies, but would in
any case only pertain to a $\sim 1/3$ fraction of minihalos.
The region of fully-molecular gas comprises a mass of $\sim 1 M_{\odot}$, with an
extent of about 100\,AU. Over a range in density, the evolution proceeds
roughly isothermally due to the
approximate balance between compressional and H$_2$ formation
heating and enhanced H$_2$ cooling. With increasing density, however, H$_2$ line
opacity, and subsequently CIE continuum opacity become important, eventually allowing
the gas to exceed $T\sim 2,000$\,K (at $n\sim 10^{16}$\,cm$^{-3}$). At this point,
H$_2$ is being collisionally dissociated, in reactions that revert the previous
three-body formation processes. In turn, the corresponding removal of the molecular
binding energy (4.48\,eV per molecule) from the thermal reservoir of the gas provides a final cooling channel in the
otherwise already optically-thick gas.

In realistic three-dimensional simulations, the emerging morphologies exhibit
complex, flattened shapes. However, the radial profiles of gas density and velocity
can approximately be described with 
spherically symmetric Larson-Penston (LP) type similarity solutions (Larson 1969, Penston 1969).
The LP solution predicts a radial density profile, $\rho\propto r^{-n_{\rho}}$, with an
exponent, $n_{\rho}=2/(2-\gamma)$, determined by the equation of state, $P\propto \rho^{\gamma}$.
For Pop~III, this was demonstrated by
Omukai and Nishi (1998) in describing their
one-dimensional radiation-hydrodynamics simulations
of primordial protostar formation. Specifically, they found $n_{\rho}\simeq 2.2$, corresponding
to an adiabatic exponent of $\gamma\simeq 1.1$, indicative of near-isothermal collapse.
Once the initial hydrostatic core has formed, however,
the self similarity is broken during the subsequent protostellar growth via accretion
(see section 3.3).

\begin{figure}
\includegraphics[width=.45\textwidth]{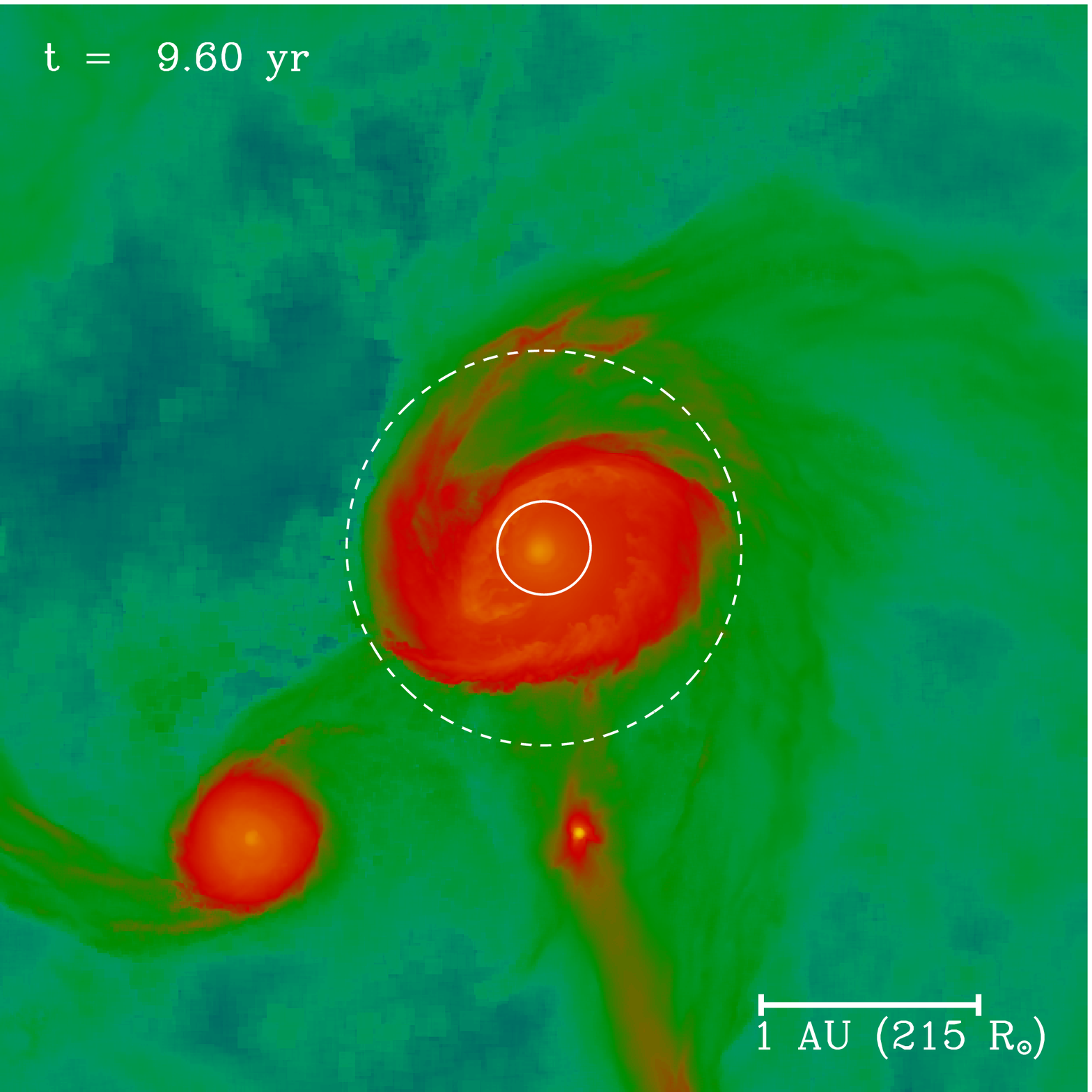}
\includegraphics[width=.45\textwidth]{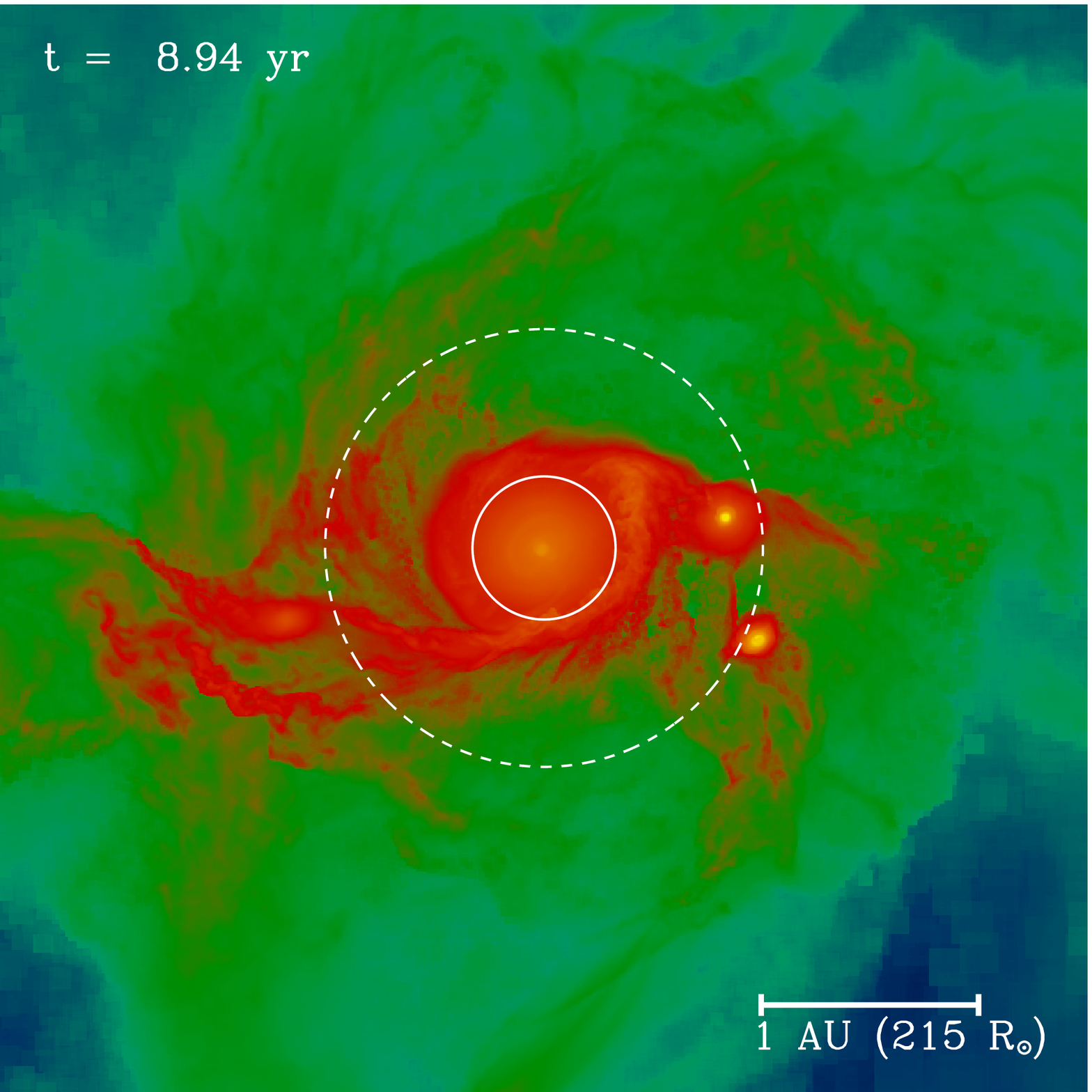}
\caption{Formation of primordial protostars. The images show the
gas density in the center of two different minihalos, with
color scale ranging from $10^{12}$~cm$^{-3}$ to $10^{21}$~cm$^{-3}$.
Each panel has a width of 5\,AU. The white circles delineate the extent of the
most massive protostar in a given minihalo, employing the photosperic ({\it dashed})
and hydrostatic ({\it solid}) criteria discussed in the text.
The diskiness and spiral structure of the accreting gas is readily apparent
(adopted from Stacy {\it et al} 2013a).
}
\label{fig3}
\end{figure}

At the end of the initial collapse, a small protostellar core has formed
at the center of the minihalo. In recent, extremely high-resolution cosmological simulations, it
has become possible to simulate the initial assembly of such a core
with unprecedented detail (Greif {\it et al} 2012, Stacy {\it et al} 2013a).
One important question in analysing these simulations is: {\it How to define the extent of a
protostar?} One possible definition is to determine the radius of the photosphere, where
$\tau \sim 1$, or within the framework of the escape probablity formalism:
$\beta_{\rm crit} = 1 - \exp (-1) \simeq 0.63$. A second definition is to determine the extent
of the region where the radial (infall) velocity is near zero, indicative of hydrostatic
equilibrium. In practice, `hydrostatic radii' are somewhat smaller than `photospheric' ones, the
latter being $R_{\rm p}\sim 100 - 200 R_{\odot}$ early on. We show some illustrative examples
in figure~3, which correspond to $\sim 10$\,yr after initial core formation. This is in accordance with
the standard theory of early protostellar structure (Stahler {\it et al} 1986), where the hydrostatic
core is hit by an accretion shock, in turn surrounded by a radiative precursor. 
The initial hydrostatic core has a mass,
$M_{\ast} \sim 10^{-2} M_{\odot}$, which is very similar to present-day, Pop~I, protostellar seeds. This
is also close to classical estimates of lower stellar masses, $M_{\rm F}$, based on opacity-limited
fragmentation (Low and Lynden-Bell 1976, Rees 1976). This near-independence of initial
protostar mass with metallicity is in accordance with theoretical expectations, where predicted
core masses only weakly depend on environmental variables, such as gas temperature (see section~3.5).
The subsequent growth of the Pop~III protostar, however, is expected
to proceed in a markedly different fashion, to be discussed next.

\subsection{Accretion and Disk Formation}

In the early Universe, protostellar accretion rates are believed to have been much larger than today, due to the higher
temperatures in the star forming clouds, which in turn is a consequence of the limited ability of the primordial gas to
cool below the $\sim 200$\,K accessible to H$_2$-cooling. This argument would still be valid, even if
a more efficient cooling agent, such as HD or metal species, were able to tie the gas temperature
to the floor set by the CMB. Minimum temperatures in star forming clouds would thus remain higher
than the canonical value in present-day molecular clouds of $\sim 10$\,K, as long as $z\gtrsim 3$:
$T_{\rm min}\simeq T_{\rm CMB}\gtrsim 11\mbox{\,K\,}(1+z)/4$ (Larson 1998, Schneider and Omukai 2010).
A useful estimate
for the protostellar accretion rate can be derived by assuming that a Jeans-mass
worth of gas collapses on its free-fall timescale (Shu 1977):
\begin{equation}
\dot{M}_{\rm acc}\simeq \frac{M_{\rm J}}{t_{\rm ff}}\simeq
\frac{c_s^3}{G}\propto
T^{3/2}\mbox{\ .}
\end{equation}
Typical accretion rates are therefore higher by two orders of magnitude in primordial star
forming regions, compared to Galactic ones (ratio $\propto(300/10)^{3/2}$).
Before 2009, it was thought that such
high accretion rates, together with the absence of dust grains
and the correspondingly reduced radiation pressure that could in principle shut off the accretion,
conspire to yield heavier final masses for Pop~III stars. Indeed, the previous `standard model',
summarized in Bromm and Larson (2004), posited that the first stars formed in isolation at
the center of a minihalo, reaching $\sim 100 M_{\odot}$, or higher (see section 3.6).

Recently, improved simulations are beginning to paint a different picture (Stacy {\it et al} 2010,
Clark {\it et al} 2011b, Greif {\it et al} 2011a, 2012). It is still
found that initially, infall is predominantly spherical, leading to the build-up of a hydrostatic
core in the center of the minihalo, with typical mass $\sim 10^{-2} M_{\odot}$ (Yoshida {\it et al}
2008). Subsequent material, however, falls in with non-negligible angular momentum, such that
stream lines do not hit the central core right away. Instead, a rotationally supported disk
is growing around the central core from the inside out. Such a disk had been suggested already
in earlier, semi-analytical studies (Tan and McKee 2004, McKee and Tan 2008). The new angle is the suggestion
that those primordial protostellar disks are ubiquitously driven towards gravitational instability.

The basic physical argument, briefly, is as follows: Because of the very high accretion rates
within a primordial pre-stellar core, $\dot{M}\lesssim 0.1 M_{\odot}$\,yr$^{-1}$, the nearly Keplerian
disk experiences rapid mass growth over a range of radii. There are strong gravitational
torques present, acting to drive mass towards the center. Even at the maximum mass transport rates that
can realistically be generated by such torques, however,
the disk cannot process the incoming material quickly enough.
Approximately, one can analyse this situation within the framework of a thin disk model, where the accretion
rate is $\dot{M}\simeq 3\pi \nu_{\rm vis} \Sigma$ (Shakura and Sunyaev 1973).
Here, $\Sigma$ is the mass per unit surface area, and the (kinematic) viscosity can dimensionally be written
in terms of the disk sound speed and pressure scale height, $H_p$, as $\nu_{\rm vis}\simeq \alpha c_s H_p$.
For gravitational torques, the Shakura--Sunyaev parameter is
$\alpha \sim 0.1 -1$. Using typical vales encountered in accretion disks around Pop~III protostars,
$\Sigma\sim 100$\,g\,cm$^{-2}$, $H_p\sim 100$\,AU\,$\sim 10^{15}$\,cm, and $c_s\simeq 10^5$\,cm\,s$^{-1}$
(Clark {\it et al} 2011b), we estimate 
$\dot{M}\lesssim 10^{-2} M_{\odot}$\,yr$^{-1}$, well below the infall rates from the envelope.

This rate imbalance will drive the disk to a state where the Toomre $Q$-criterion
for global gravitational stability, $Q=c_s\kappa/(\pi G \Sigma)>1$, is violated. Here,
$\kappa$ is the epicyclic frequency, equal to the angular velocity ($\Omega$) in a Keplerian disk.
Thus, the disk is subject to global perturbations,
such as spiral modes. To enable fragmentation, however, a second, stronger, criterion needs to
be considered. This is the Gammie criterion (Gammie 2001), stating that for a density
perturbation to survive the disruptive effect of disk shear, thus enabling successful
fragmentation, the cooling timescale has to be shorter than the orbital one: $t_{\rm cool}<
3 \Omega^{-1}$. The simulations have yielded the surprising result that disk temperatures
remain close to $T\simeq 2,000$\,K, even at densities high enough for opacity effects to
become important (see section 3.1.2). There is an effective thermostat provided by
the collisional dissociation of molecular hydrogen, which absorbs 4.48\,eV per dissociation
event. This is somewhat of a `knife-edge' effect, however. If the disk were just a bit hotter,
resulting from an internal or external heating source, it may be possible to stabilize the disk accretion
mode, thus again leading to a single star at the center. Including the luminosity generated by, possibly
highly time-variable,
gas accretion onto the protostar(s) does somewhat delay fragmentation, but cannot suppress it
completely (Smith {\it et al} 2011, 2012a). This result, however, does only pertain to the early
evolution, where photo-ionization has not yet become important (see section~3.6).
The behavior of the Pop~III disks follows the same trends, in terms of fragmentation
and stability, that govern protostellar disks in general (Kratter {\it et al} 2010).

The simulations mentioned above were carried out with SPH
or with the new moving-mesh code AREPO (Springel 2010). Recent adaptive-mesh refinement (AMR)
simulations of the disk build-up have presented a study of the resolution-dependence involved
(Turk {\it et al} 2012). Those authors caution that at their highest resolution, corresponding
to 64 cells per local Jeans length, or higher (Truelove {\it et al} 1998), the emergence
of the disk is suppressed by small-scale, subsonic turbulence, or at least delayed. The
latter qualification arises from the fact that the AMR study does not employ sink particles,
and therefore cannot follow the disk evolution for much beyond the initial hydrostatic core
formation. It is thus still somewhat of an open question how robust the claimed
inevitable drive towards disk instability is. However, the latest AREPO simulation by
Greif {\it et al} (2012) reaches an effective resolution that matches, or exceeds, the
Turk {\it et al} (2012) requirement, and it does confirm the emergence of a gravitationally
unstable disk. 

What is the likely outcome of this disk fragmentation mode? We will discuss this
in the next 4 subsections.

\subsection{Multiplicity}

All simulations that employed a technique allowing to go beyond the initial core formation event,
either sink particles (Stacy {\it et al} 2010, Clark {\it et al} 2011b, Greif {\it et al} 2011a), or
the stiffening of the equation of state due to large opacities (Greif {\it et al} 2012), have shown
that the disk fragments into a small multiple, often dominated by a binary. Strong hints for binarity
were already seen in the simulation by Turk {\it et al} (2009). In their case, however, binarity was
a rare event, as it arose through the near-simultaneous collapse of neighboring density peaks. Fragmentation
was thus induced during the initial collapse phase, where most of the time secondary density peaks are
only temporary and do not survive, rendering this fragmentation mode unlikely. This is to be compared
with the near-ubiquity of the disk fragmentation mode, which happens after the initial collapse has ended.
Pop~III binarity was also seen in idealized numerical experiments, where primordial gas was endowed
with different amounts of initial rotational and thermal energy (Machida {\it et al} 2008a). Those experiments
established that binarity for the first stars was at least plausible, but the interpretation
was subject to uncertainties related to the unrealistic, non-cosmological initial conditions chosen.

Thus, to a reasonable level of confidence, we have an existence proof that the first stars typically formed as members of a small multiple system. As a next step, we now wish
to establish the statistical properties of the resulting systems, such as binary
mass ratios, orbital parameters, and overall binary fraction. All of these key quantities are not yet known with any certainty.
In a recent, medium-resolution, suite of simulations, the evolution of Pop~III stellar systems has been
traced for sufficiently long ($\sim 5,000$\,yr) to begin to constrain their vital statistics (Stacy and Bromm 2013b).
Specifically, any given Pop~III star has a $\sim 50$\% chance of being accompanied by a companion, with a
nearly log-normal distribution of binary periods, peaking at $\sim 900$\,yr. Again, those results
are uncertain, and are not fully converged spatially and temporally due to the considerable computational
expense involved. Future work with more efficient codes, run on even more powerful computers,
is clearly needed.

\subsection{Lower Mass Limit}
The classical theory of placing a lower limit to the mass of (sub-) stellar objects is
opacity-limited fragmentation (Low and Lynden-Bell 1976, Rees 1976). The idea is that
the Jeans mass continues to decrease during (near-) isothermal collapse, until the
density becomes high enough to render the gas opaque, so that the compressional 
heating can no longer be radiated away; afterwards, the Jeans mass increases again
with density. The minimum fragment mass can be estimated in a robust way where
all the details of radiative emission processes and transport are absorbed into an efficiency
factor $f=f(Z)\lesssim 1$, depending only weakly on metallicity, $Z$, and other factors,
such as collapse geometry and clumpiness of the medium:
\begin{equation}
M_{\rm F}\simeq M_{\rm Ch} f^{-1/2} \left(
\frac{k_{\rm B}T}{m_{\rm H}c^2}
\right)^{1/4}
\mbox{\ ,}
\end{equation}
where $M_{\rm Ch}$ is the Chandrasekhar mass.
With $f\sim 1$, indicating an emission process close to a blackbody, as is appropriate for
highly opaque gas, and $k_{\rm B}T\sim 1$\,eV for $\sim 10^4$\,K protostellar gas, one finds
$M_{\rm F}\simeq 10^{-2} M_{\odot}$, close to what is found in simulations.
The key point is that this limit only very weakly depends on environment and metallicity, 
$M_{\rm F}\propto f^{-1/2} T^{1/4}$, such that the same lower mass limit should apply
to both Pop~I and Pop~III stars. An independent confirmation of this near-universality
of the opacity limit has been found in numerical experiments, where gas collapse was
investigated with different metallicities (Omukai 2000).
Uehara {\it et al} (1996) considered the opacity limit within the framework of
an idealized cylindrical collapse model. They did, however, identify the moment of
fragmentation with the density where the primordial gas becomes opaque to
H$_2$ line cooling. Since this occurs at moderate densities, and does not take into
account CIE cooling, which remains optically thin to much higher densities (see section~3.1),
their estimate for the lower Pop~III mass limit was correspondingly larger,
$M_{\rm F}\simeq M_{\rm Ch}\simeq 1 M_{\odot}$. 

The precise value of the lower-mass limit to Pop~III stars is important in predicting
whether any truly metal-free stars could have survived until the present day. To
have survived for the entire history of the Universe, a star needs to have a mass
$\lesssim 0.8 M_{\odot}$. Some recent simulations (Clark {\it et al} 2011b, Greif {it et al} 2011a)
show that a number of fragments, represented by sink particles, are ejected from the cloud center
though N-body dynamics. These `run-aways' could be candidates for Pop~III survivors, provided
that the results are confirmed in future simulations that can realistically model protostar-protostar 
close encounters. For the latter, it is important to include the possibility that orbital kinetic energy
is dissipated through tidal forces, or that protostars are dragged to the center through
viscous forces, where they are merged with one another. This behavior has been found
in radiation-hydrodynamical simulations of present-day massive star formation (Krumholz {\it et al} 2009),
and has been seen in the recent AREPO calculations that do not employ sink particles to
represent protostars (Greif {\it et al} 2012). Even if low-mass Pop~III stars once formed, there is
still an obstacle to identify them in surveys of metal-poor stars in the Galaxy (Frebel {\it et al} 2009).
This is the `pollution limit', set by the slow accretion of ISM material onto the surfaces
of metal-poor stars (Yoshii 1981, Iben 1983). Pop~III survivors could thus be masqueraded as
extreme Pop~II stars. If low-mass Pop~III stars were able to trigger a low-level wind, accretion from
the ISM could be prevented, thus preserving their pristine surface composition (Johnson and
Khochfar 2011). The question, however, is whether such a wind could be maintained.

Note however, that such masked Pop~III stars should reflect the average
chemical abundance pattern of the Galactic ISM. It is important to realize that there is
currently no indication for such a population of stars. A much more efficient pollution
channel would become available in a sufficiently close Pop~III binary system, where the
surviving secondary could have accreted material from an evolved asymptotic giant branch (AGB)
companion star (Suda {\it et al} 2004, 2013). Such binary `self-enrichment' could readily account for
light elements (C, N), but may need some fine tuning for the heavier ones. Regarding this scenario,
it is crucial to carry out long-term monitoring of the most metal-poor stars to detect any signs
of binarity in their lightcurves.

The current null result 
in finding any metal-free stars anywhere in the Milky Way provides strong empirical evidence
that Pop~III stars typically were more massive than $\sim 1 M_{\odot}$. The question of
the lower mass limit for Pop~III stars would then largely be academic, and the important
problem is to identify their typical, characteristic mass (see section~3.7).

\subsection{Upper Mass Limit}

To first order, we can estimate the upper mass limit of a Pop~III star by considering the
asymptotic growth, as follows: $M_{\ast,{\rm up}}\sim \dot{M}_{\rm acc} t_{\rm acc}$,
where $t_{\rm acc}$ is the effective accretion timescale. For massive stars, one often equates
this with the Kelvin-Helmholtz timescale, $t_{\rm KH}\sim 10^5$\,yr, which is the time
needed for a star to reach the main-sequence (MS). Further assuming a time-averaged accretion
rate in the primordial pre-stellar core of $\dot{M}_{\rm acc}\sim 10^{-3} M_{\odot}$\,yr$^{-1}$ (Tan and McKee 2004), we obtain:
$M_{\ast,{\rm up}}\sim 100 M_{\odot}$. An even more robust limit can be derived by
setting:
$t_{\rm acc}\sim t_{\ast}$, where $t_{\ast}\simeq 3$\,Myr is the MS lifetime of a very massive
star. Evaluating the integral:  
$M_{\ast,{\rm up}}\simeq \int_0^{t_{\ast}}\dot{M}_{\rm acc} dt$, typically results in limits of
$M_{\ast,{\rm up}}\simeq 500-600 M_{\odot}$ 
(Abel {\it et al} 2002, Omukai and Palla 2001, 2003, Bromm and Loeb 2004).

\begin{figure}
\includegraphics[width=.9\textwidth]{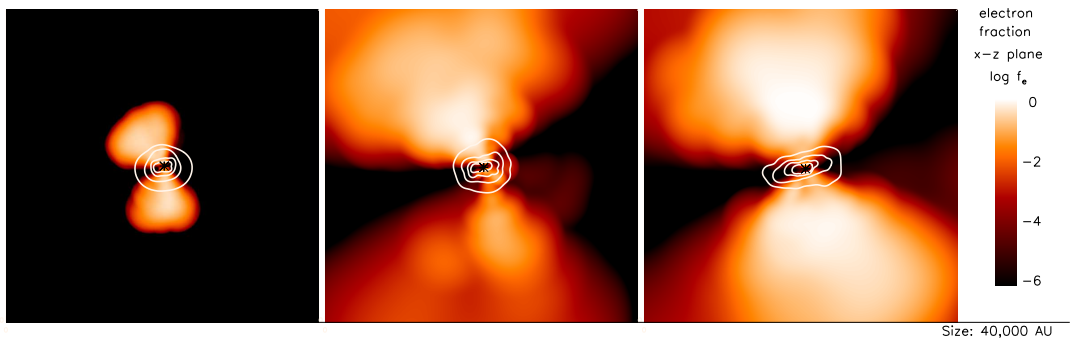}
\caption{Protostellar radiative feedback. Shown is the ionization fraction of the gas at 1500, 2000,
and 3000\,yr after the initial sink formed. It can be seen that an ultracompact H\,{\small II} region is
emerging, centered on the most massive protostar, and exhibiting a characteristic hour-glass shape.
The box size is $40,000$\,AU. The white density contours delineate the density structure of the disk,
with contours ranging from $10^{7.5}$\,cm$^{-3}$ to
$10^{9}$\,cm$^{-3}$. The H\,{\small II} region gradually grows and dissipates the disk from above and from below.
Eventually, this feedback will shut off accretion, thus limiting the mass of the star
(adopted from Stacy {\it et al} 2012b).
}
\label{fig4}
\end{figure}

Although pedagogically instructive, this simple picture has to be refined. Even if one ignores the
competition from multiple stars (see section 3.4) in accreting the available gas, the key physics in 
setting the upper mass limit for the first stars is radiative feedback from the growing protostar.
This feedback cannot be neglected once a protostar has grown to about $10 M_{\odot}$ (Omukai and Inutsuka 2002).
Within the context
of a semi-analytical model, McKee and Tan (2008) have identified the photo-evaporation of the disk, where
a bipolar ultra-compact H\,{\small II} region surrounding the central protostar `boils away' the disk. Once the
disk has thus been destroyed, accretion has to stop, and the Pop~III star has reached its final mass,
in this case typically:
$M_{\ast,{\rm up}}\simeq 140 M_{\odot}$. Recently, first attempts have been made to
carry out
fully self-consistent radiation-hydrodynamical simulations of the Pop~III accretion process
(Hosokawa {\it et al} 2011, 2012, Stacy {\it et al} 2012b). These simulations confirm
the basic picture suggested in McKee and Tan (2008), but revise the upper mass limit to somewhat
lower values:
$M_{\ast,{\rm up}}\simeq 30-60 M_{\odot}$. Since star formation inherently involves randomness,
it is likely that in rare cases, even higher masses can be assembled. This leads us to next consider
the distribution of Pop~III stellar masses, the initial mass function (IMF) of the first stars. 

It is important to keep in mind that models can be constructed where Pop~III stars reach 
masses that are significantly higher,
$M_{\ast,{\rm up}}\gtrsim 10^3 M_{\odot}$, possibly even becoming `supermassive' 
($M_{\ast,{\rm up}}\gtrsim 10^6 M_{\odot}$). In each case, the physics or initial
conditions invoked are somewhat exotic, implying that any very massive Pop~III objects
were quite rare. Such scenarios may involve dark stars (discusses in section~4.2), or
direct collapse pathways, where primordial clouds can avoid fragmentation, forming
very massive objects in the centers of DM halos (see Bromm and Yoshida 2011 
and references therein, and more recently Latif {\it et al} 2013b). In the latter case, the
DM halos need to be massive enough to reach virial temperatures $T_{\rm vir}\gtrsim
10^4$\,K (see section~2), to enable near-isothermal collapse at gas temperatures
$T\sim T_{\rm vir}$. An intriguing sub-class of the direct collapse models suggests
an intermediate stage of very massive `quasi-stars', where the `star' is powered
by the accretion of a massive envelope onto a central BH seed (Begelman {\it et al} 2008).
The formation of such rare, very massive objects raises the tantalizing prospect of
extremely energetic explosions at the edge of the Universe, possibly involving general 
relativistic instabilities (Johnson {\it et al} 2013).

\subsection{The Primordial IMF}

The ultimate goal of a theory of Pop~III star formation would be to predict the primordial IMF
in a completely ab-initio fashion. Despite significant progress made, we are far away
from this utopian goal. In general,
the stellar IMF is a complicated function of mass, but it is often convenient
to simply write it as a power law, valid for a given mass range. Specifically,
one considers the number of stars per unit mass:
\begin{equation}
\frac{dN}{dM_{\ast}}\propto M_{\ast}^{-x}\mbox{\ ,}
\end{equation}
where the present-day IMF is characterized by the famous Salpeter slope of $x= 2.35$ (Salpeter 1955).
To understand what the {\it typical} outcome of the star-formation process is, one
can ask: {\it Where does most of the available mass go?} 
This can be calculated as follows:
\begin{equation}
\bar{M_{\ast}}\simeq\frac{1}{N_{\ast}} \int_{M_{\ast,{\rm low}}}^{M_{\ast,{\rm up}}}M_{\ast}\frac{dN}{dM_{\ast}}dM_{\ast}
\sim 4 \times M_{\ast,{\rm low}}
\end{equation}
where we have used the Salpeter value for $x$,
and $M_{\ast,{\rm low}}$ and $M_{\ast{\rm up}}$ are the lower and upper mass limits, respectively.
Furthermore,
$N_{\ast}\simeq \int (dN/dM_{\ast})dM_{\ast}$
is the total number of stars in the newly formed cluster.
In general, the lower mass limit sets the typical stellar scale,
as long as $x>2$.

\begin{figure}
\includegraphics[width=.9\textwidth]{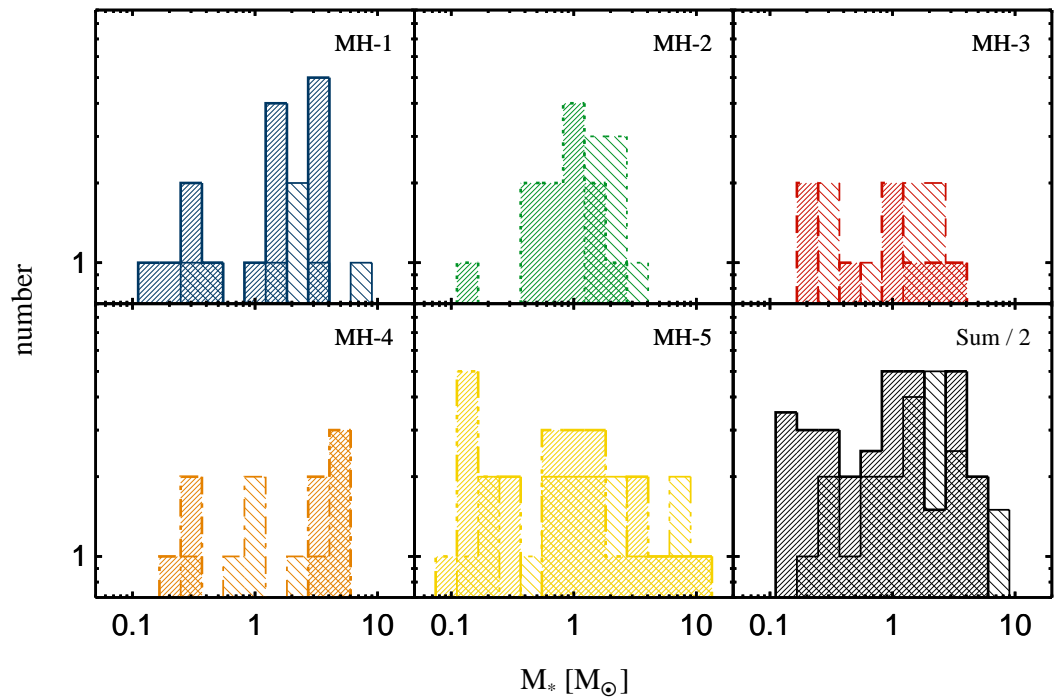}
\caption{Early Pop~III mass function. Shown is the situation after the first $1,000$\,yr of continued accretion, fragmentation and merging of sink particles, for the 5 independent
minihalos considered. The lower-right panel depicts the cumulative, average mass function, renormalized
for convenience. The latter, extending from $\sim 0.1 - 10 M_{\odot}$, 
is nearly flat, indicating a top-heavy distribution, where most of the mass is locked up in fragments at
the upper end. In each panel, the result for two different assumptions on how readily sink particles
can merge upon close encounters are included: no merging of sink particles at all ({\it dark shading}),
and aggressive merging ({\it light shades}). It is evident that the distributions are shifted toward
higher mass in the latter case
(adopted from Greif {\it et al} 2011a).
}
\label{fig5}
\end{figure}

The Pop~I IMF is thus bottom-heavy, with
$\bar{M_{\ast}}\sim 0.5 M_{\odot}$. What is the situation for Pop~III?
Based on the arguments presented above there is a consensus that the first star IMF was
top-heavy, but possibly in a less extreme fashion than what was thought pre-2009. It is
not yet possible to predict the Pop~III IMF in any detail, but recent simulations are
beginning to give us hints. The Greif {\it et al} (2011a) simulation was able to
investigate the early phases of Pop~III star formation, following the evolution for
$\sim 1,000$\,yr after the initial core was formed. Due to the efficiency of AREPO, it
was also possible to address cosmic variance by considering a number of minihalos that
were statistically independent. The average mass function, at $t\sim 1,000$\,yr, is 
nearly flat, corresponding to $x\simeq 0$ (figure~5). Within the framework introduced above, this
would correspond to a top-heavy mass function. The caveat here is that there is still
the crucial extrapolation involved getting one from the $\sim 1,000$\,yr that {\it can}
be simulated in an ab initio fashion, to the $\sim 10^5$\,yr that the accretion process
is likely to last (see section~3.6). Taking the current hints at face value, one can
summarize the situation as follows:
The first stars had a characteristic (typical)
mass of $\bar{M_{\ast}}\sim \mbox{\, a few\,}\times 10 M_{\odot}$, and a power-law slope that
may have been shallower than Salpeter. The latter assumption
is quite uncertain, however.

\subsection{Stellar Rotation}

In general, next to mass, stellar rotation is the second most important parameter
in determining a star's evolution, nucleosynthesis, and final fate
(Langer 2012, Maeder and Meynet 2012). For Pop~III, very little had
been known regarding rotation, and only recently has this important
subject become accessible to realistic simulations, tracing the build-up
of stellar rotation all the way from cosmological initial conditions
(Stacy {\it et al} 2011b, 2013).
The highest-resolution study indicates that the protostars quickly
develop a roughly solid-body rotation profile, while their
surface rotation velocities range from $\sim 80 - 100$\% of the Keplerian velocity, $v_{\rm Kep}$
(see figure~6). The caveat here is that the extremely high-resolution simulations have followed
the protostellar assembly process
for a mere $\sim 10$\,yr. This limitation is due to the sheer computational expense of carrying out
simulations that do not employ sink particles.

\begin{figure}
\includegraphics{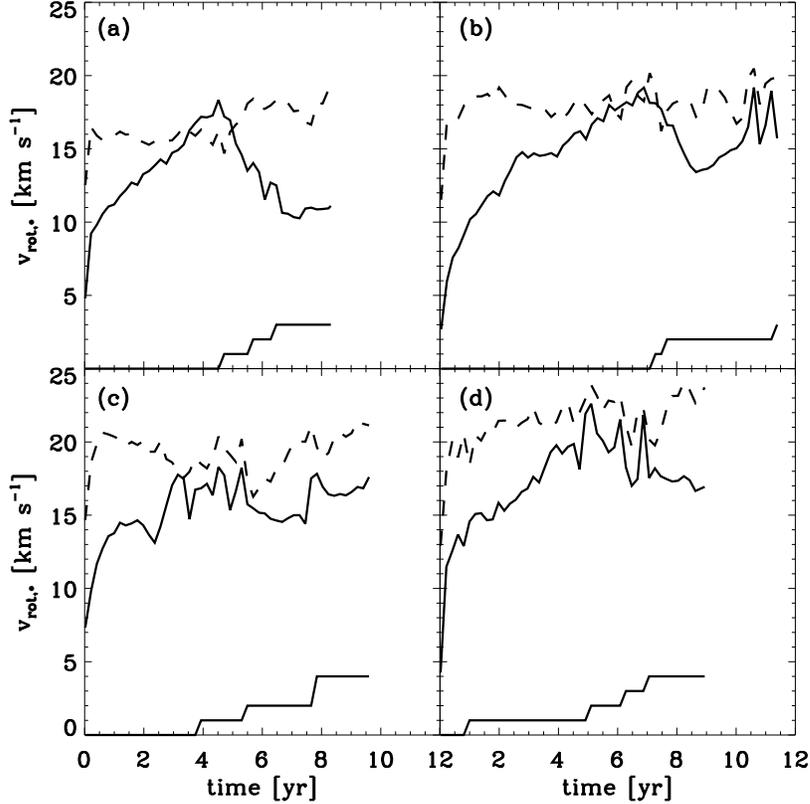}
\caption{Pop~III protostellar rotation. {\it Solid lines:} Surface rotation rates, evaluated at
the equator vs. time. {\it Dashed lines:} The Keplerian velocity vs. time. The solid lines at the
bottom of each panel indicate the number of mergers undergone by the most massive protostar in
a given halo. It is evident that rotation velocities are a sizable fraction of the Keplerian, or break-up,
speed. Note that the brief period in panel (a), where the velocity becomes larger than the break-up
value, is an artefact of the azimuthal averaging procedure
(adopted from Stacy {\it et al} 2013).
}
\label{fig6}
\end{figure}

High rotation velocities persist even after undergoing
multiple merger events, indicating that these occur in a prograde fashion.
There is also little
evidence of any correlation between the large-scale properties of each host
minihalo, in particular its spin, and the angular momentum of its largest protostar or the total
number of protostars formed in the minihalo. The decoupling between large-scale DM environment and
the small-scale pre-stellar cores, however, may not be so surprising, if one considers
the rapid redistribution of angular momentum in the protostellar accretion disk
due to gravitational torques and hydrodynamical shocks.
The statistics here is, however, still quite limited, as
Stacy {\it et al} (2013) only consider 10 minihalos. A different conclusion was reached by de Souza {\it et al} (2013),
who argue that overall DM halo properties are reflected in the final outcome of star formation, specifically
in shaping the Pop~III IMF. Their simulation, however, had limited resolution, so that an analytical recipe
had to be used to model disk fragmentation and the build-up of a stellar system.

The key challenge for future work is to follow the build-up of protostellar angular momentum
for significantly longer timescales.
As the protostars continue to grow and continue on
to the MS, rotation could alter the protostar’s life in a number
of ways (Maeder and Meynet 2012). Rotation rates which persist at sufficiently high
levels, for instance, may allow for mass loss through stellar
winds generated at the so-called $\Omega\Gamma$-limit, where centrifugal forces near the equator ($\Omega$)
assist continuum-driven radiation pressure near the Eddington limit ($\Gamma$).
Substantial mass loss could then result, even in the absence of any
line-driven winds (Kudritzki
2002). This would reduce the final mass of the star and thus
may alter its fate encountered upon death.
Metal production during the
lifetime of the star would also be enhanced in general
(Ekstr\"{o}m {\it et al} 2008, Yoon {\it et al} 2012). Stellar temperature
and luminosity will be modified as well, an effect that is possibly
greatly enhanced if rotational mixing is sufficient for
the star to undergo chemically-homogeneous evolution (CHE), though not all
studies agree that CHE can take place in rotating Pop~III
stars (Ekstr\"{o}m {\it et al} 2008). CHE may furthermore
provide a mechanism for a Pop~III star to evolve into a Wolf-Rayet (WR)
star and eventually a GRB without being in a tight binary
(Yoon and Langer 2005, Woosley and Heger 2006, Yoon {\it et al}
2012). Finally, CHE may also lower the minimum
mass at which a star will undergo a pair-instability supernova (PISN) death
from 140 to $\sim 65 M_{\odot}$ (Chatzopoulos and Wheeler 2012,
Yoon {\it et al} 2012). Such a PISN is an extremely energetic explosion, that
completely disrupts the progenitor star, leaving no remnant behind (further discussed
in section~6.1).

Observations of the pattern of chemical elements in the atmospheres of metal-poor stars
(see section~6.3) can provide constraints on the
rotation rate of the first stellar generations. A tantalizing example is provided 
by the recent analysis of the abundance pattern in red giant stars within the possibly
oldest Globular Cluster in the Galactic bulge (Chiappini {\it et al} 2011). These
stars show, puzzlingly, the {\it simultaneous} presence of elements typically associated
with high-mass, explosive, nucleosynthesis (r-process), and intermediate-mass, AGB-type,
enrichment (s-process). Models of massive, rapidly-rotating, metal-poor stars suggest
that such hybrid nucleosynthesis may originate in a single stellar source. This is
indirect evidence in support of the hypothesis that the first stars were typically rapid
rotators. It is important, however, to emphasize that these are early days in constraining
the rotation state of Pop~III stars.

\section{Beyond the Standard Model}
\subsection{Magnetic Fields}

An important missing ingredient of high-$z$ structure formation is a comprehensive
understanding of the first magnetic fields (Rees 2000, Widrow {\it et al} 2012). Specifically: How were
primordial seed fields generated? How and when were they amplified to dynamically
significant strengths? And what was their impact on the formation of the first
stars, galaxies, and quasars? It is well-established that magnetic fields are a key ingredient
in the physics of present-day star formation (Stahler and Palla 2004, McKee and Ostriker 2007). They
play a key role in shaping the (MHD-) turbulence in the molecular birth clouds, in the transport of
angular momentum, and in limiting the efficiency of the star formation process. It is likely that such
effects were important in primordial star formation as well, and recent work has begun to take them
into account.

\begin{figure}
\includegraphics[width=0.9\textwidth]{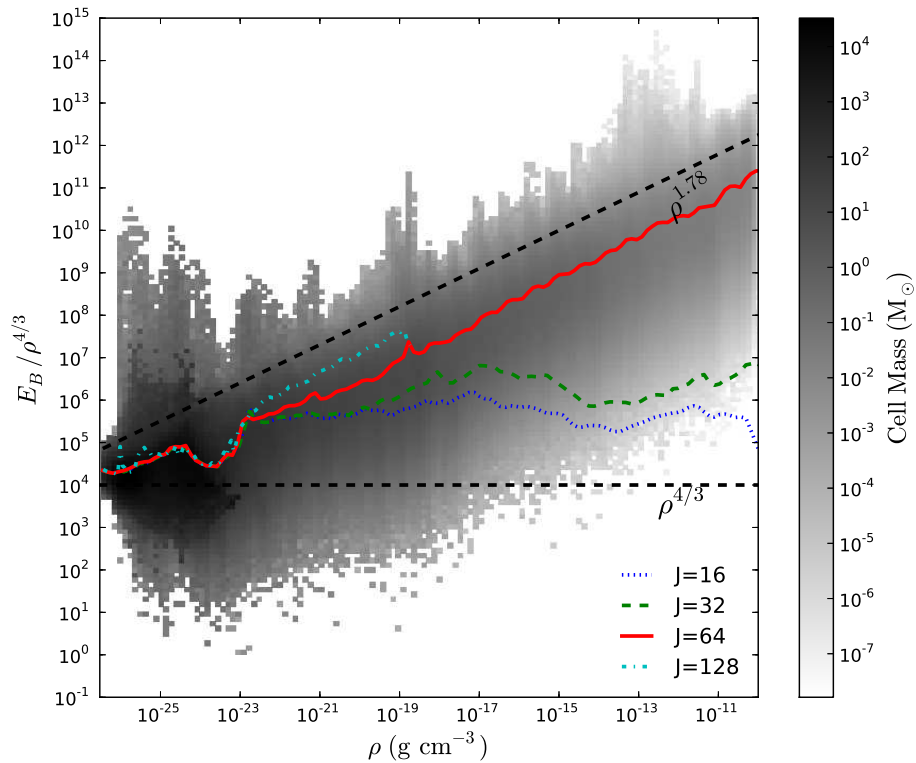}
\caption{Dynamo action during minihalo collapse. Shown is the magnetic energy density, $u_{B}=B^2/8\pi$, scaled
to the value expected for flux freezing only ($u_{B}\propto \rho^{4/3}$), vs. gas density. The colored lines
show the situation for four different resolutions, measured in cells-per-Jeans-length ($J$). As is evident,
the magnetic field is amplified above the flux-freezing baseline in each case, but convergence occurs only
for $J\gtrsim 64$ ({\it red line}). The grey-shaded pixels indicate the mass per cell for a given $u_{B}-\rho$ bin,
for the $J=64$ case
(adopted from Turk {\it et al} 2012).
}
\label{fig7}
\end{figure}

Seed fields in the high-$z$ IGM could arise either from exotic processes in the very early Universe,
possibly related to electro-weak or QCD phase transitions or to inflation (reviewed in Widrow {\it et al} 2012), or
through the action of the Biermann battery in collapsed structures (Biermann 1950). The latter mechanism has been
studied with cosmological simulations of minihalo collapse, indicating that field strengths of
$B\lesssim 10^{-17}$\,G can be generated on scales of $\sim 100$\,pc (Xu {\it et al} 2008, Doi and Susa 2011).
Such fields in themselves would not suffice to influence the subsequent collapse and the detailed process
of star formation, even when allowing for the amplification resulting from ideal-MHD flux freezing, $B\propto
\rho^{2/3}$, where $\rho$ is the gas density.
Further efficient amplification of any seed field through the action of a kinematic dynamo, where frozen-in field lines
are repeatedly folded by a velocity field that is endowed with net helicity, is therefore required (Kulsrud 2005).
The small-scale turbulence generated during the virialization of the minihalo, and subsequently through
the ongoing infall of material onto the central core, has been shown to efficiently feed such a dynamo 
(Schleicher {\it et al} 2010, Sur {\it et al} 2010, Schober {\it et al} 2012), possibly driving field strengths
to equipartition with the energy of the random, turbulent motion.
Recently, it has become feasible to address the dynamo amplification process with
ab initio cosmological simulations, at least during the initial stages of collapse (Turk {\it et al} 2012, Latif
{\it et al} 2013a). A key result is that very high resolution, in excess of 64 cells per Jeans length
in the language of the Truelove criterion discussed above (see figure~7), is needed to properly resolve 
the turbulence and the magnetic field saturation (see also Federrath {\it et al} 2011).

What are the implications if dynamically significant magnetic fields were indeed present in the disks
surrounding growing Pop~III protostars? Critical field levels for triggering the magneto-rotational
instability (MRI) would likely be reached, thus providing an additional, strong mechanism to
transport angular momentum (Tan and Blackman 2004). In addition, conditions may well be in place
to launch MHD jets, leading to the magneto-centrifugal removal of mass and spin (Machida {\it et al} 2006, 2008b).
Finally, the disks could be heated through ambipolar diffusion and the dissipation of the
MHD turbulence on the viscous scale, thus suppressing fragmentation. The net effect on the resulting mass
spectrum is not clear, however, and needs to be elucidated with future dedicated simulations.

\subsection{Self-annihilating Dark Matter}

In the standard model of first star formation, the DM component only plays a passive
role by providing the gravitational potential wells, where gas dissipation can take
place, and where turbulent motions are generated. Within the popular class of
weakly interacting massive particle (WIMP) candidates, however, a more direct effect
could arise (see Bertone {\it et al} 2005 for a review). Since (Majorana) WIMPs are their own antiparticles,
they could self-annihilate provided that the DM density is sufficiently high. The complex decay channels
would eventually give rise to normal particles, such as electron-positron pairs and photons, 
which in turn would heat and ionize the gas. Stars could thus be stabilized by a non-nuclear source
of energy, before the very high densities for nuclear burning are reached. The resulting red supergiant
stars, with typical radii of $\sim 1$\,AU, have been termed `dark stars' (Spolyar {\it et al} 2008,
Freese {\it et al} 2008), although this is somewhat of a misnomer. These stars are not `dark' at all,
but instead possess normal stellar photospheres, albeit at much lower effective temperatures
compared to the standard massive Pop~III stars (Freese {\it et al} 2008, Yoon {\it et al} 2008, Natarajan {\it et al} 2009, Hirano {\it et al} 2011).
A further consequence of the red dark star colors, implying the virtual absence of ionizing UV photons, is the
significantly reduced protostellar feedback, possibly allowing the star to grow to very large masses,
$M_{\ast}\gtrsim 10^{6} M_{\odot}$ (Freese {\it et al} 2010).
Such supermassive stars would then also be
extremely bright, bringing them within reach of detection with the JWST (Zackrisson {\it et al} 2010, Ilie {\it et al} 2012). One key strength of
the dark star proposal is that the DM self-annihilation rate is fixed by the
CDM freeze-out density, $<\sigma v>\simeq 3\times 10^{-26}$\,cm$^{3}$\,s$^{-1}$, which in turn is
of the expected order for weakly interacting particles. Indeed, this is the essence of the famous
`WIMP miracle' (Bertone {\it et al} 2005).

Vigorous follow-up work has fleshed out many of the missing physical ingredients.
An important point was made by Ripamonti {\it et al} (2010), who
calculated the initial protostellar collapse phase with one-dimensional simulations. They conclude
that DM-annihilation (DMA) heating is insufficient to arrest the collapse towards protostellar densities.
The momentum of the infalling material pushes the object beyond the possible bottleneck, where 
DMA heating temporarily balances gas cooling. This cooling bottleneck was initially interpreted
as a sufficient criterion to stabilize the star at relatively low densities. DMA heating could still
play a role, by providing a long-lived source of additional energy in the center of a Pop~III star,
thus significantly prolonging their otherwise very short lifetime (Iocco {\it et al} 2008). Such late-time replenishment
of the central DM density could result from baryon-WIMP scattering capture, provided that the 
relevant cross sections are high enough (Iocco 2008, Sivertsson and Gondolo 2011).

\begin{figure}
\includegraphics[width=0.8\textwidth]{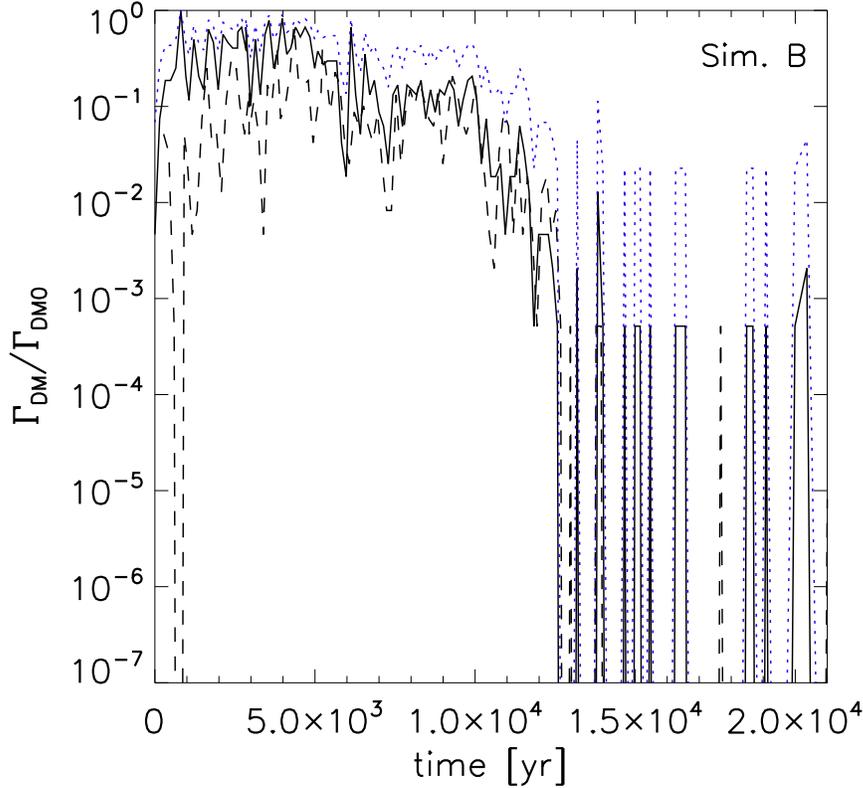}
\caption{Suppression of DM heating through Pop~III multiples. Shown are the DM heating rates,
relative to the initial, maximum values, evaluated at the locations of the two most massive
protostars, represented by sink particles in the simulation (solid and dashed black lines).
The decline in the relative rate for the capture of DM particles through
WIMP-baryon scatterings is plotted, as well (dotted blue line). It is evident that
the presence of the Pop~III star-disk system steadily suppresses the initially
high DM density, modulated by an oscillatory pattern
(adopted from Stacy {\it et al} 2012a).
}
\label{fig8}
\end{figure}

Recent debate has focused on whether the conditions for dark star formation were ever realized in
a realistic cosmological setting. A possible weakness of the dark star scenario is that
a high degree of symmetry in the dark matter density profile seems to be required. In one-dimensional
models of Pop~III star formation in the center of a minihalo, such symmetry is fulfilled by design;
the growing protostar is located right on top of the central DM cusp, thus being subject to
maximal DMA heating. Within the new paradigm of multiple Pop~III stars that arise through 
gravitational instability in protostellar disks (see section~3.3), the required symmetry
may be broken, however (Stacy {\it et al} 2012a). Indeed, the Pop~III multiple lessens the
impact of DMA heating by first displacing the accreting protostars from the central cusp, and,
secondly, by reducing the central DM density through gravitational scattering interactions
(see figure~8).
The caveat in the Stacy {\it et al} (2012a) study is that the presence of a Pop~III multiple
was pre-supposed, arguing that DMA heating would only become important at even higher densities,
thus not affecting the disk fragmentation process. It is possible, however, that
DMA heating could stabilize the protostellar disk, preventing any fragmentation (Smith {\it et al} 2012b);
any suppression of the DM density via gravitational N-body dynamics would then not take place.
The latter simulation has assumed a fixed DM potential, and it is not clear whether a more
realistic `live' DM halo would behave in the same way. The debate is ongoing, and it may
be a while until we have reached a more complete understanding.

\subsection{Cosmic Rays}

The presence of cosmic rays (CRs) is well known to significantly influence the 
ISM in local galaxies, in terms of heating and ionizing deeply embedded
gas clouds (see Tielens 2005 for an overview). One can speculate that the same
may well be the case for Pop~III star formation. The main challenge is to
predict the strength of any CR background, together with its
energy distribution, at high redshifts. Starting with the premise that CRs can
be produced in the wake of magnetized SN explosions due to Fermi acceleration (see
Schlickeiser 2002 for a pedagogic introduction), attempts have been made to construct
the build-up of such a CR background, linked to the high-$z$ star formation rate
density (Jasche {\it et al} 2007, Stacy and Bromm 2007). A qualitatively different
production channel for ultra-high energy CRs is the decay of exotic particles that
may have survived from the very early Universe (Shchekinov and Vasiliev 2004, Ripamonti
{\it et al} 2007). Such CRs would then interact with
the CMB, giving rise to energetic, hydrogen and helium ionizing photons.
Given the huge uncertainties in arriving at these predictions, a promising cross-check
is provided by linking any early CR background to the abundance of $^6$Li observed
in the atmospheres of metal-poor Galactic halo stars (Rollinde {\it et al} 2005, 2006, Asplund {\it et al} 2006).
A pre-galactic CR spallation channel could result in a $^6$Li bedrock abundance, parallel to the famous
Spite plateau for $^7$Li (Spite and Spite 1982). The latter is taken as indicative for Big Bang nucleosynthesis,
but the $^6$Li plateau appears to exceed any Big Bang production by about a factor of 1000. To explain a near
plateau, however, there must have existed a production mechanism that predated the bulk of star formation.
The Pop~III scenario would nicely fit those requirements, such that a pervasive CR background produced in the wake
of the first SNe could be responsible for establishing the $^6$Li plateau abundance.
The problem with such a scenario is, however, that the
concomitant CR heating might result in unrealistically high IGM temperatures.

\begin{figure}
\includegraphics[width=0.8\textwidth]{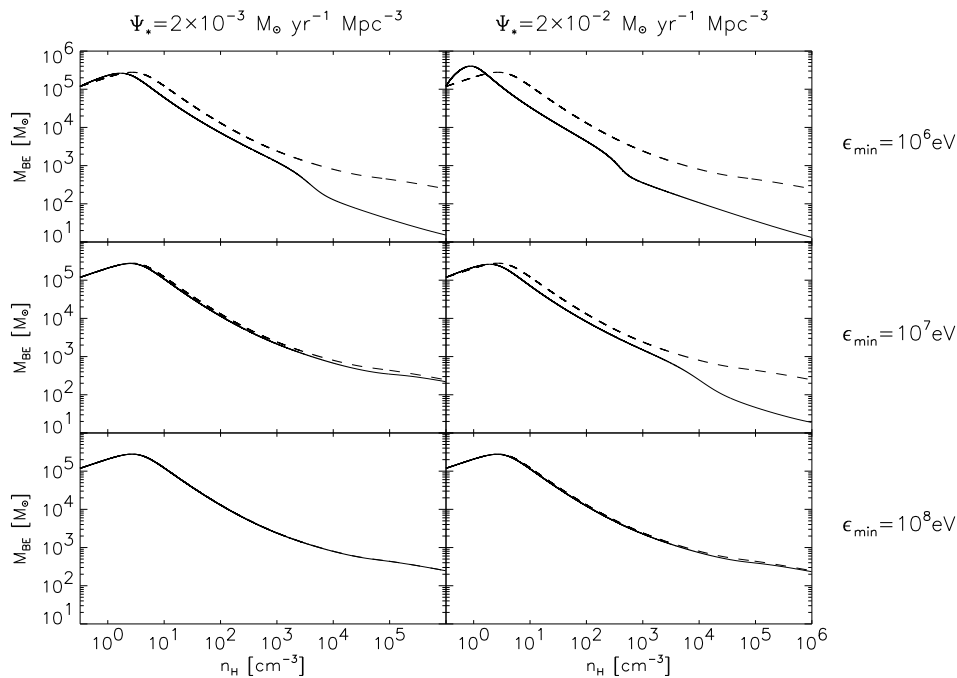}
\caption{Impact of a CR background on the Pop~III fragmentation mass. Shown is 
the Bonnor-Ebert mass, closely related to the Jeans mass and providing a rough
estimate for the stellar mass, as a function of hydrogen number density. Here, 
primordial gas is collapsing into minihalos at $z\sim 20$, suffused in a
CR background of different amplitude and minimum energy. The former is related to
the Pop~III star formation rate density, labelled on the top, and the latter is
marked on the right. In each panel, the solid line gives the evolution with a
CR background present, and the dashed line that without. It can be seen that
the Pop~III mass is reduced by a factor of $\sim 10$, provided that a sufficient
flux of low-energy CRs is present
(adopted from Stacy and Bromm 2007).
}
\label{fig9}
\end{figure}

In principle, a sufficiently high CR background could modify the thermal history of primordial
gas, collapsing into minihalos, in important ways. Specifically, if the CR hydrogen ionization rate
exceeds $\zeta_{\rm CR}\gtrsim 10^{-19}$\,s$^{-1}$, the additional abundance of free electrons
is able to activate the HD cooling channel, which in turn can tie the gas temparature to
that of the CMB (see section~3.1).
This is in contrast to the canonical, H$_2$ moderated collapse, where temperatures
never drop below $\sim 100$~K. The lower temperature may be reflected in correspondingly
decreased Jeans masses, possibly resulting in lower-mass stars, of order
$\sim 10 M_{\odot}$ (see figure~9). These estimates, however, are very uncertain, relying on simple analytical 
arguments; it will be important to firm them up with high-resolution numerical simulations, similar
to what can now be done in the classical Pop~III case. The impact of CR heating and ionization
is likely much less important when primordial gas collapses into more massive, $M_h \gtrsim 10^8 M_{\odot}$, host halos (Stacy and 
Bromm 2007). The reason is that the boost in (collisional) ionization in the stronger virialization shocks  
is already capable of activating the HD channel, thus establishing the minimum allowed temperature (Greif {\it et al} 2008);
any additional CR ionization would thus be ineffective. An interesting angle to this dynamic
has recently been added by Inayoshi and Omukai (2011), investigating the impact of CR ionization
on collapsing metal-free gas that is subject to a strong soft UV, Lyman-Werner (LW), background.
Such LW photons are able to photo-dissociate molecular hydrogen.
It has thus been suggested that LW irradiated primordial gas may avoid ever forming H$_2$
molecules, which would suppress cooling of the gas, thus preventing fragmentation and star formation (Haiman
{\it et al} 1997). The atomic gas would then engage in a near-isothermal collapse, possibly
leading to the direct formation of a supermassive black hole (Bromm and Loeb 2003a, Volonteri and Bellovary 2012).
The argument by Inayoshi and Omukai (2011) posits that the same stellar population that would have
established the LW background also might have produced a CR, and possibly X-ray, background. The CR
background, in turn, would counter-act the LW suppression effect, due to the boost in free electron
catalysts, thus promoting the abundance of H$_2$. It is currently not clear what the net effect of
these coupled backgrounds will be, warranting further detailed studies.

\section{Second Generation Stars}
Within our current understanding (summarized in setion~3), metal-free star formation
in minihalos constitutes a somewhat singular case, with the initial conditions given
by $\Lambda$CDM cosmology, and resulting in a top-heavy mode of star formation.
This is in stark contrast with subsequent star formation, as observed locally and inferred
by observations of the highest-redshift galaxies currently accessible to our most powerful
telescopes. Here, initial conditions are largely decoupled from the dark matter structure
of the large-scale Universe, and the observed outcome is dominated by `normal', typically
low-mass stars. An important problem, therefore, is to understand how and when cosmic star
formation did transition from the early, high-mass dominated mode to the more normal one
later on. Theoretical models have proposed two qualitatively different, second-generation
transition populations: stars that form already out of, at least slightly, enriched gas,
produced and dispersed by Pop~III SN explosions (Pop~II stars); and stars that form out
of still metal-free gas, but in environments that have been modified in important ways, 
compared to the simple initial conditions in the canonical minihalo case. Such environmental
complexity could arise in a number of ways, e.g., through the presence of ionizing radiation
backgrounds, established by previous star formation, or through collissional ionization in 
the wake of strong shocks when more massive halos undergo virialization. This latter population
has been termed `Pop~III.2', to indicate that it still involves {\it metal-free} gas, but
to differentiate it from the classical, minihalo population, now more precisely termed
`Pop~III.1' (O'Shea {\it et al} 2008, McKee and Tan 2008, Bromm {\it et al} 2009).
We will next discuss these transitional populations in turn, followed by an assessment of how
the emergence of supersonic turbulence may fundamentally change the character of early
star formation.

\subsection{Radiative Feedback (Pop~III.2)}

The basic idea behind the formation of Pop~III.2 stars is to channel the primordial gas,
prior to the onset of gravitational instability, through a phase of significantly increased
ionization (Johnson and Bromm 2006). The over-abundance of free electrons can then boost
the production of H$_2$, thus lowering the temperature to the point where HD cooling
can kick in, finally enabling the gas to reach the temperature floor set by the
CMB. Such lowered temperatures may result in smaller fragment masses, given the 
scaling of the Jeans mass: 
$M_{\rm J}\propto T^{3/2}$. Given the requirement of an increased abundance of
free electrons, there are qualitatively different pathways towards Pop~III.2, and we
will here briefly discuss the most plausible ones.

The first pathway arises from photo-ionization in the neighborhood of massive Pop~III stars.
A possible environment is provided by
relic H\,{\small II} regions, around central Pop~III stars
that have just died (Yoshida {\it et al} 2007). The ensuing non-equilibrium recombination typically
results in a boosted abundance of H$_2$ and HD, thus allowing cooling to the CMB floor.
The problem for triggering second-generation star formation, however, is the low density encountered
in such environments, as a consequence of outflows driven by the photo-ionization heating (Kitayama {\it et al} 2004, Whalen {\it et al} 2004, Alvarez {\it et al} 2006). A second
pathway, relying on collisional ionization in strong virialization shocks, may be more promising
(Greif {\it et al} 2008). Here, the idea is that metal-free gas collapses into DM halos that are
more massive than minihalos, exhibiting virial temperatures of $T_{\rm vir}\gtrsim 10^4$\,K, what are
often termed `atomic cooling halos' due to their ability to cool through the emission
of Lyman-$\alpha$ photons (see figure~10). The key challenge for this scenario is how to realistically keep the gas
metal-free, given that star formation would normally already occur in the progenitor minihalos within
the merger tree of the atomic cooling halos (Johnson {\it et al} 2008). 

\begin{figure}
\includegraphics[width=0.7\textwidth]{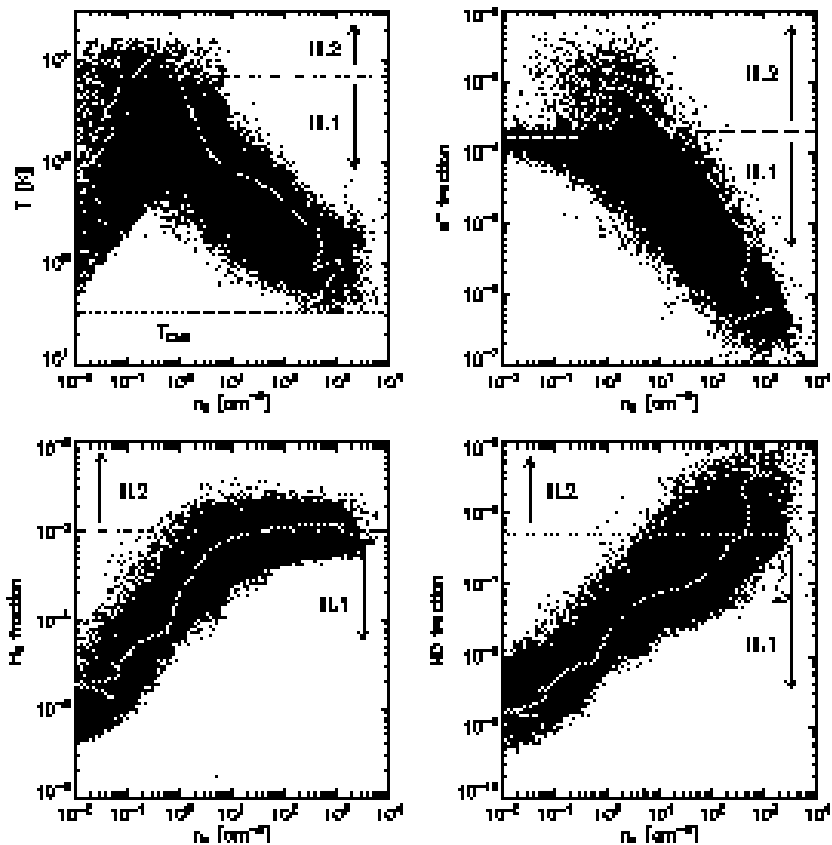}
\caption{Enhanced cooling in pre-ionized primordial gas. Shown are the gas
temperature, free electron fraction, HD fraction, and H$_2$ fraction
as a function of hydrogen number density, for primordial gas collapsing
into an atomic cooling halo at $z=10$. Here, the boost
in (collisional) ionization at the virialization shock enables additional
H$_2$ molecules to form, in turn leading to an enhanced abundance of HD.
The HD cooling channel is able to cool the gas to the temperature of the CMB
at these redshifts. Under these conditions, Pop~III.2 stars are predicted
to form. The solid lines, superimposed on the black dots, trace the path of
a typical fluid element that follows the HD cooling channel. In addition,
the arrows denote regions in $T-n$ phase space that separate the Pop~III.1
and III.2 cases, such that only gas that traverses through the `III.2' region
can successfully cool to the CMB
(adopted from Greif {\it et al} 2008).
}
\label{fig10}
\end{figure}

Again, there are two principal ways to prevent pre-enrichment of the gas. First, {\it all} Pop~III.1 stars
that had formed in the progenitor minhalos could have directly collapsed into black holes, without
any concomitant dispersal of metals (see Karlsson {\it et al} 2013). Since there are of order ten such
progenitors (Greif {\it et al} 2008, Wise and Abel 2008), the likelihood for such complete
`sterilization' is small. In rare cases, the infall of the primordial gas into an atomic cooling halo
could thus proceed in a `cold-collapse' mode, which is similar to the cold-accretion mode seen 
in simulations of the formation of larger galaxies later in cosmic history (Birnboim and Dekel 2003,
Kere\v{s} {\it et al} 2005).
Under such conditions, the Pop~III.2 channel is indeed activated (Greif {\it et al} 2008). Existing 
simulations, however, have not yet followed the protostellar collapse and accretion to sufficiently high densities,
and for sufficiently long after initial core formation, to explore the resulting stellar mass spectrum. In
particular, it is not clear whether the analytical prediction that the typical mass of Pop~III.2 stars
is lower than for Pop~III.1 is correct (see Clark {\it et al} 2011a).

There is a second way to keep the gas metal-free, prior to the emergence of atomic cooling halos, by preventing
any Pop~III star formation in the minihalo progenitors altogether. This can occur in the presence of a modest
LW radiation background, which would act to photo-dissociate H$_2$, thus depriving the primordial
gas of its only viable low-temperature coolant. Once gas infall is triggered by Lyman-$\alpha$ cooling
in the deeper gravitational potential wells of an atomic cooling halo, densities will eventually
rise to the point where self-shielding becomes important (Wolcott-Green and Haiman 2011). Molecules can then
form, allowing gas temperatures to reach $T\sim 200-300$\,K. Recent simulations, starting from realistic
cosmological initial conditions, have shown, however, that within this `hot collapse' mode, the HD cooling
channel is never activated (Wolcott-Green {\it et al} 2011, Safranek-Shrader {\it et al} 2012). Indeed, the subsequent thermal
evolution is very similar to the canonical Pop~III.1 minihalo case. The reason that here HD cooling
never becomes important is that H$_2$ formation is delayed to densities that are high enough to enable
self-shielding. At these high densities, the free electron abundance is already quite low, due to
the prevelance of recombinations, with their $\propto n^2$ scaling. This in turn prevents the boost
in H$_2$, which otherwise would similarly have boosted the HD abundance and allowed the temperature to
drop below $\sim 100$\,K (see also Wolcott-Green and Haiman 2011).

\subsection{Chemical Feedback: Critical Metallicity (Pop~II)}

Once the first heavy elements have been produced and dispersed in energetic Pop~III SN explosions,
the physics and chemistry of subsequent star formation will be fundamendally changed.
Atomic and ionic metal species, as well as dust grains, will provide efficient coolants, establishing
conditions similar to the present-day Milky Way ISM. Grain-catalyzed molecule formation may in 
addition greatly impact the chemical make-up of the star forming clouds. It had been argued early on,
that there may exist a threshold enrichment level, governing the transition between the top-heavy Pop~III
and more normal Pop~II mode (Omukai 2000). Indeed, this has given rise to
the concept of a `critical metallicity', such that low-mass (Pop~II) star formation is enabled for
$Z\gtrsim Z_{\rm crit}$ (Bromm {\it et al} 2001, Schneider {\it et al} 2002). The question then is: What
is its value, and is it a universal constant or does it show a complex dependence on environment? A different
question is whether metallicity is the only important variable in driving the Pop~III -- Pop~II transition,
or whether other factors, such as redshift or magnetic field level plays a
role as well. The redshift dependence would come in via a decreasing temperature floor set by the CMB (Larson 1998),
and a `critical $B-$field' through a threshold to enable the MRI in the 
protostellar disk (Silk and Langer 2006), or through similar MHD effects (see section~4.1). E.g., with saturation-strength fields present, magnetic pressure
driven outflows may limit the accretion efficiency, thus resulting in less massive stars (Tan and Blackman 2004).

Assuming that metallicity is the key ingredient in enabling low-mass star formation, two classes of models have 
been suggested, leading to different values of the critical metallicity. The first class suggests that cooling
due to to fine-structure lines from atomic or ionic metal species allows low-mass stars to form, leading to
typical values of $Z_{\rm crit}\simeq 10^{-3.5} Z_{\odot}$ (Bromm {\it et al} 2001, Smith {\it et al} 2009,
Safranek-Shrader {\it et al} 2010). More specifically, lines due to C\,{\small II} and O\,{\small I} have been
identified as dominant coolants (Bromm and Loeb 2003b), with a possible role for 
Fe\,{\small II} and Si\,{\small II} as well (Santoro and Shull 2006). The second class proposes cooling due
to dust grains, synthesized in the first SNe (Gall {\it et al} 2011), as key drivers in enabling low-mass star formation
(Schneider et al 2006, Schneider and Omukai 2010). Within the dust-cooling model, the key parameter is
a critical dust-to-gas ratio, $D_{\rm crit}\simeq 4\times 10^{-9}$, which can be re-written as a criterion
on the minimum required dust depletion factor at any given level of metallicity: $f_{\rm dep}\gtrsim D_{\rm crit}/Z$
(Schneider {\it et al} 2012a). Evidently, dust-cooling models can accommodate conditions of extremely low-$Z$
in high-redshift star forming clouds, and still enable fragmentation. Somewhat idealized simulations, where
the equation of state was pre-computed and not self-consistently coupled to the thermal evolution of the
collapsing cloud, have confirmed the analytical predictions for dust-induced fragmentation (Tsuribe and 
Omukai 2006, Clark {\it et al} 2008).

\begin{figure}
\includegraphics[width=0.7\textwidth]{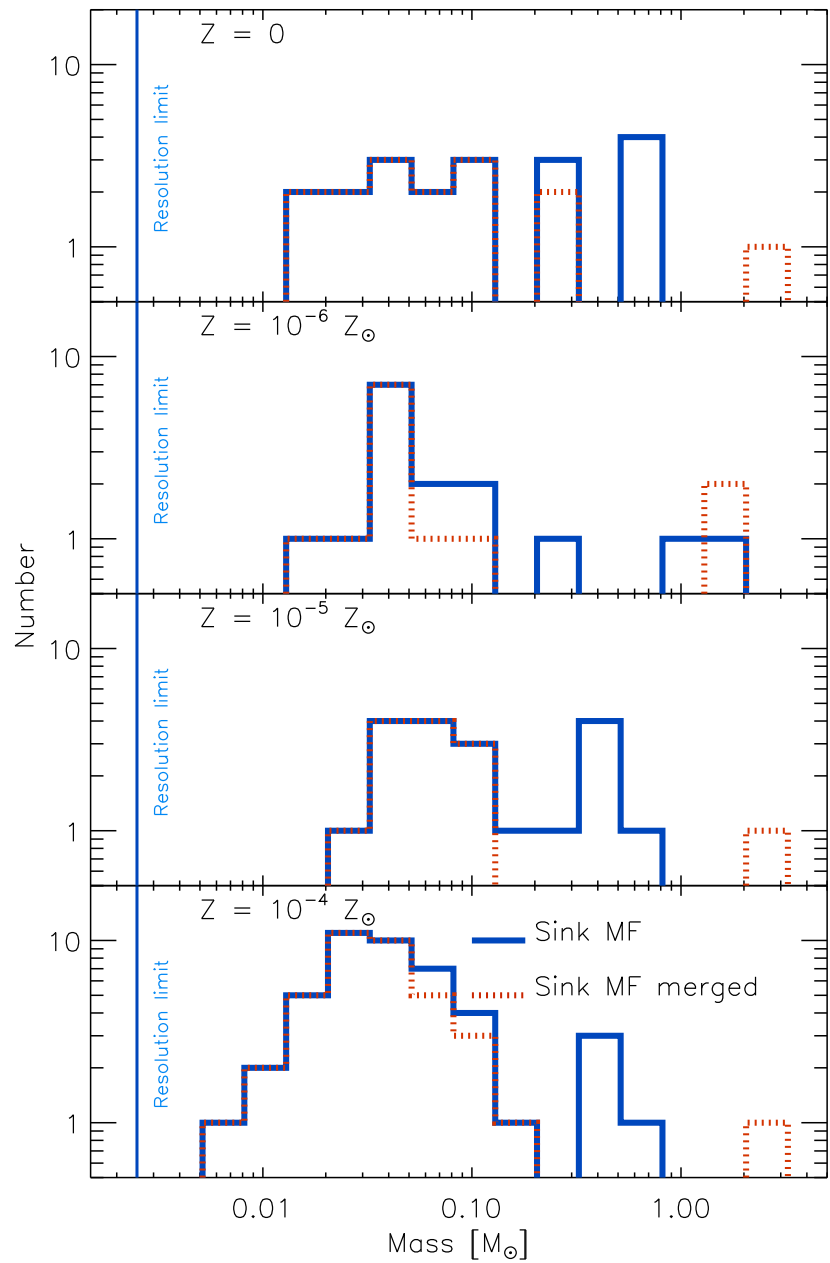}
\caption{Sink mass function at low metallicities. The sink masses
are proxies for protostellar masses. There appears to be a qualitative
change at $Z\lesssim 10^{-5}Z_{\odot}$, where the mass spectrum flattens,
thus leading to a top-heavy situation. The vertical lines indicate the
(mass) resolution limit of the simulations. The red histograms show
the results for a different implementation of sink particles, allowing
them to merge upon close encounter; this, however, does not change the
basic result
(adopted from Dopcke {\it et al} 2013).
}
\label{fig11}
\end{figure}

A series of numerical investigations has suggested that, indeed, there is no critical metallicity at all,
and that gas fragmentation always extends to low masses, even for $Z\simeq 0$ (Jappsen {\it et al} 2009a,b, Dopcke {\it et al} 2013).
Jappsen {\it et al} (2009b) have argued that the fine-structure threshold at $Z\simeq 10^{-3.5} Z_{\odot}$
disappears if H$_2$ molecules are present as well. However,
in the Bromm {\it et al} (2001) setup, H$_2$ was deliberately neglected, because the pre-existing stars
necessary to produce the first metals would presumably also establish a pervasive LW radiation background;
the LW photo-dissociation would then destroy all H$_2$, at least in the low-column-density minihalos where
self-shielding cannot yet act. Thus, the critical metallicity would indeed exist. Recently, Dopcke
{\it et al} (2013) have fixed the problem with `hard wiring' the equation of state, and now calculate the
dynamics, chemistry and thermal evolution self-consistently. They find the intriguing result that,
although there again seems to be no critical metallicity below which the gas could no longer fragment into very low-mass objects,
the IMF nevertheless becomes more top-heavy for $Z\lesssim 10^{-5} Z_{\odot}$ (see figure~11). The reason, however, is that
the mass spectrum flattens, such that the relative importance of the high-mass end increases (see section~3.7).

The physics behind the Pop~III--Pop~II transition is evidently very complex. To really make progress here,
guidance from observations is needed. We will briefly review the relevant lessons from stellar archaeology
below (section~6.3).

\subsection{Turbulence and the First Clusters}

\begin{figure}
\includegraphics[width=.9\textwidth]{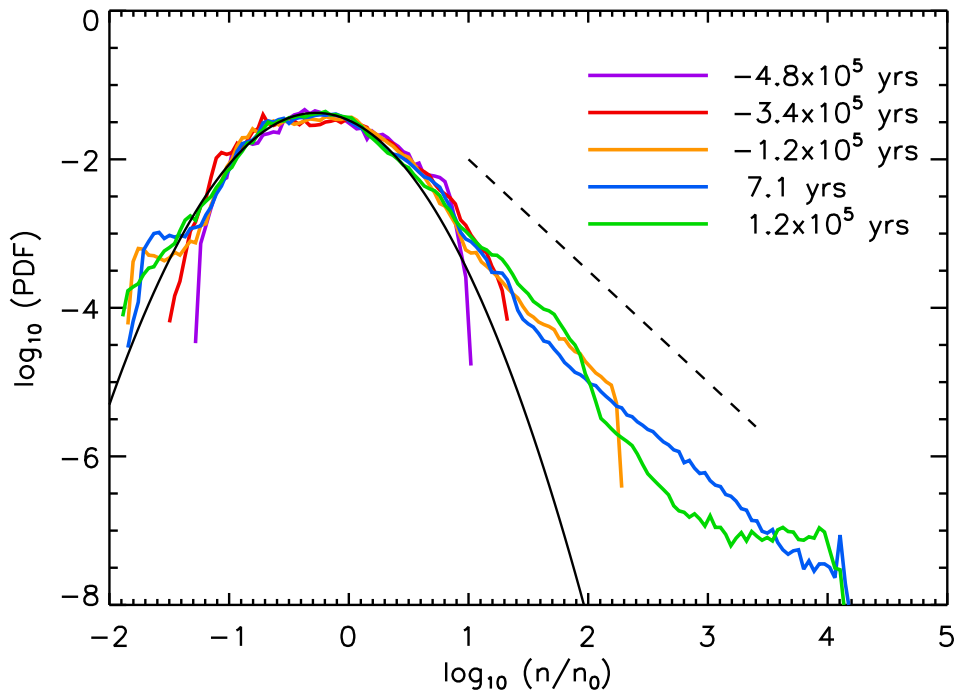}
\caption{Density fluctuations in the center of the first galaxies.
The presence of supersonic turbulence is manifested in the characteristic log-normal
probability distribution. At late times, the effect of self-gravity imprints a power-law
extension towards the highest densities. It is possible that the turbulently structured gas
will give rise to a high-mass slope in the stellar IMF similar to the present-day Salpeter one
(adopted from Safranek-Shrader {\it et al} 2012).
}
\label{fig12}
\end{figure}

There may be another, qualitatively very different, aspect to the Pop~III -- Pop~II transition, related to
the character of turbulence in the star forming material.
Whereas the first stars form in minihalos, where turbulence is subsonic, or at most mildly transonic, due
to the shallow gravitational potential wells involved, the emergence
of the first galaxies marks the onset of supersonic
turbulence. More specifically, simulations have determined that this transition occurs at the scale of
an atomic cooling halo, about 100 times more massive than a minihalo (Wise and Abel 2007, Greif {\it et al} 2008).
The Reynolds number in the center of atomic cooling halos at $z\gtrsim 10$ is indeed very large,
$Re\sim 10^9$, indicating a highly-turbulent situation, and the Mach number,
$Ma\sim V/c_s\sim v_{\rm vir}/c_s\sim 10$, indicates supersonic flows. The
last estimate assumes a virial velocity typical for an atomic cooling
halo, $v_{\rm vir}\sim 10 \mbox{\,km\,s}^{-1}$, and
the sound-speed of H$_2$-cooled gas, $c_s\sim
1$\,km\,s$^{-1}$. Note that a similar estimate for a minihalo would give
$Ma\sim 1$. The presence of supersonic turbulence is expected
to have important consequences for high-redshift star formation, similar to the situation in the present-day Universe
(Larson 2003, Mac~Low and Klessen 2004, McKee and Ostriker 2007). Among them are a clustered mode of star
formation, and the role of self-similarity in shaping the power-law extension of the IMF toward the high-mass end. 

In general, supersonic turbulence generates density fluctuations in the star-forming gas cloud.
Statistically,
these can be described with a log-normal probability density function (PDF):
\begin{equation}
f(x)dx=\frac{1}{\sqrt{2\pi \sigma^2_x}}\exp\left[-\frac{(x-\mu_x)^2}{2\sigma^2_x}
\right]dx\mbox{\ ,}
\end{equation}
where $x\equiv\ln(\rho/\bar{\rho})$, and $\mu_x$ and $\sigma^2_x$ are the
mean and dispersion of the distribution, respectively. The latter two are
connected: $\mu_x=-\sigma^2_x/2$.
Numerical simulations have shown that the dispersion of the density
PDF is connected to the Mach number of the flow (see McKee and Ostriker 2007): $\sigma^2_x\simeq
\ln(1+0.25Ma^2)$. Inside the first galaxies, one finds values close
to $\sigma_x\simeq 1$ (Safranek-Shrader {\it et al} 2012). Similar to the well-studied case of
isothermal, supersonic turbulence (Kritsuk {\it et al} 2011), the central gas in the first galaxies
exhibits the imprint of self-gravity: a power-law
tail toward the highest densities, on top of the log-normal PDF at
lower densities, which is generated by purely hydrodynamical effects (figure~12).

A useful way to characterize turbulence is by way of velocity structure functions. Prieto {\it et al} (2011)
have measured the second-order function, $S_2 \propto \ell^{\zeta(2)}$, in cosmological AMR simulations of 
atomic cooling halo collapse. Here, $\ell$ is the distance over which the velocity differences are evaluated,
and $\zeta(2)\simeq 1.04$, as determined in the simulation, over the range $\sim 100-500$\,pc.
One then finds a velocity-size relation, $S_2^{1/2}\propto
\ell^{0.52}$, which is very similar to the `Larson-law' velocity-size relation for present-day molecular clouds
(Larson 1981). The huge Reynolds number of the underlying turbulence implies a very large dynamic range between
the feeding scale, roughly the virial radius of an atomic cooling halo, $R_{\rm vir}\sim 1$\,kpc, and the
viscous dissipation scale. Even the most highly-resolved simulations to date, employing some form of
adaptive spatial refinement, will, therefore, not be able to fully capture the turbulent flows encountered.
First attempts have been made to incorporate such unresolved turbulence with subgrid-scale (SGS) modeling,
properly matched onto the resolved large-eddy simulation (Latif {\it et al} 2013a). It is not yet clear how
robust such SGS modeling is, and how appropriate the calibrations used. Overall, these are still early
days in the study of the coupled gravito-turbulent star formation process in the high-redshift Universe, but
it is likely that we will see rapid progress. A prime challenge is to work out the character of clustered
star formation in the first galaxies, which must carry the imprint of the supersonic turbulence availabe
inside of them.

\section{Empirical Signatures}

To make progress, we need to calibrate and test our increasingly sophisticated numerical
simulations with observational constraints. The upcoming suite of next-generation facilities,
in particular the JWST and the ground-based extremely large telescopes, promises to do just that.
However, despite their
exquisite sensitivity at near-IR wavelengths, even these observatories may not be able to directly
probe the first stars, unless they formed in massive clusters (Pawlik {\it et al} 2011), or were gravitationally lensed (Rydberg {\it et al} 2012). The only opportunity to probe
individual Pop~III stars may be to catch them at the moment of their explosive death. This could involve
extremely energetic SN events, such as hypernovae or pair-instability SNe (Hummel {\it et al} 2012, Pan {\it et al} 2012), or GRBs. The latter fate depends on whether Pop~III stars could give 
rise to suitable collapsar progenitors, involving rapidly rotating massive stars (MacFadyen and Woosley 1999).
Since Pop~III stars are predicted to fulfill both requirements (as discussed in section~3), GRBs are
likely to occur out to the highest redshifts. We will discuss these explosive deaths first, in terms of
their in-situ observability `one star at a time', followed by a nicely complementary probe in our immediate
cosmic neighborhood: the abundance patterns of chemical elements detected in the most metal-poor stars in
our Milky Way and the Local Group. This approach, sometimes termed stellar archaeology or near-field
cosmology (Freeman and Bland-Hawthorn 2002),
places indirect constraints on the properties, in particular the masses, of the first SNe.

\subsection{Supernovae: direct detection}

\begin{figure}
\includegraphics[width=.9\textwidth]{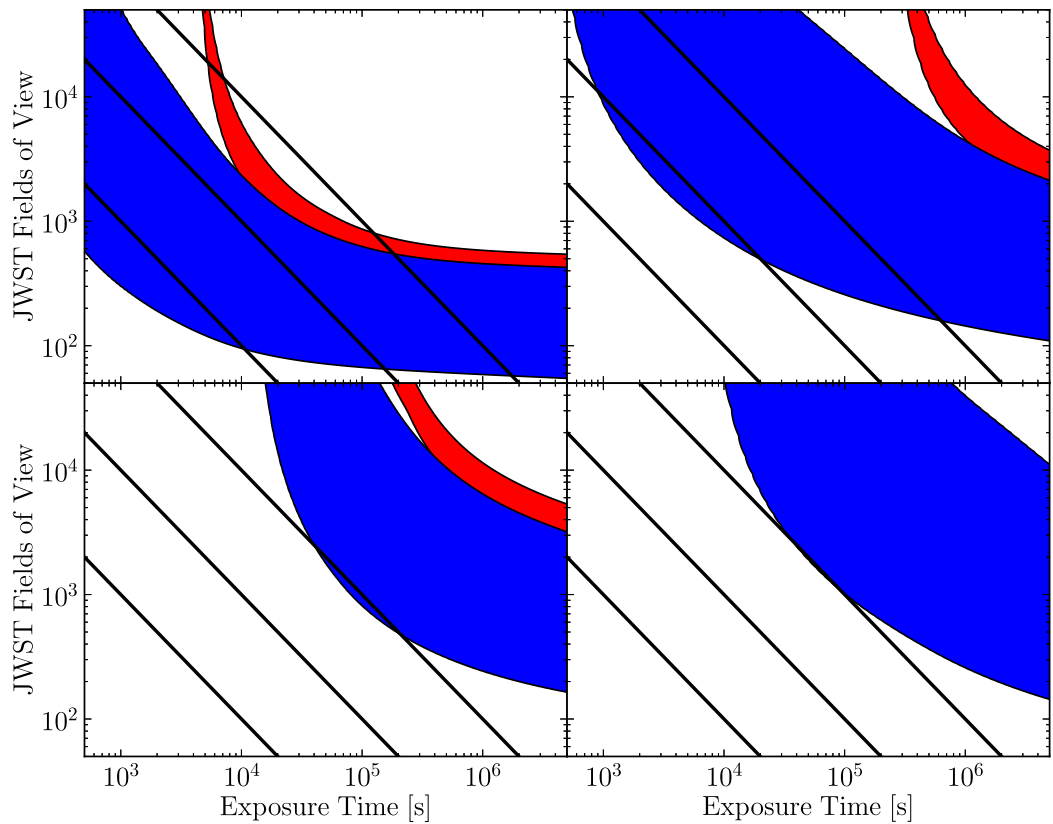}
\caption{Observability of primordial PISNe with the JWST, specifically its NIRCam F444W filter.
The panels show results for different Pop~III progenitor
models, from a $250 M_{\odot}$ red supergiant model (top-left), to one with 
$175 M_{\odot}$ (bottom-right). The other two panels depict intermediate cases, in terms of ease
of observability. The blue regions show the required combination of exposure time and number of
JWST field-of-views, in order to detect about 10 sources. The red strips indicate the same
information for $z>15$ PISNe only, where detection becomes increasingly difficult.
The black lines represent the total number of pointings in a given campaign of total duration
$10^6, 10^7$, and $10^8$\,s. It can be seen that the optimal survey strategy involves modestly
deep ($t_{\rm exp}\gtrsim 10^4$\,s) exposures, combined with a wide-area mosaic
(adopted from Hummel {\it et al} 2012).
}
\label{fig13}
\end{figure}

Probing the complete functional form of the Pop~III IMF will remain out of reach, even for the JWST. It
is, therefore, important to constrain this distribution from the extremes, the high- and low-mass ends.
At the low-mass end, one can conduct surveys for Pop~III survivors in our Milky Way, which would imply
the existence of stars with $M_{\ast}\lesssim 0.8 M_{\odot}$ (discussed in section~6.3).
For very high masses, roughly in the range $\sim 140 - 260 M_{\odot}$, the first stars are predicted
to die as extremely energetic PISNe. Those events are bright enough to be picked up by the JWST photometrically
out to very high $z$ (Kasen {\it et al} 2011, Dessart {\it et al} 2013, Whalen {\it et al} 2013a). However, there are two principal
problems in hunting down any Pop~III PISNe. The first is that although PISNe are very bright, they are also very rare
(Hummel {\it et al} 2012). Going deeper, to JWST Near Infrared Camera (NIRCam)
exposures of $t_{\rm exp}\gtrsim 10^4$\,s, would not deliver any further
sources; instead, the optimal search strategy is assembling a wide mosaic of modest exposure each (see figure~13).
The second observational challenge is to `type' a PISN (Pan {\it et al} 2012). As their
lightcurves are quite extended, with rest-frame plateaus lasting for over a year, the effect of cosmological
time dilation would stretch this to over a decade in the observer frame, for sources at $z\gtrsim 10$.
Thus, a high-redshift PISN would not appear as a clearly identifiable transient in most cases, and the
standard `point-and-repeat' selection techniques, that are so successful at $z\sim 1$ for Type~Ia SNe, would
not work. There should, however, be a detectable time-dependence in the photometry, towards increasingly
redder colors, and possibly in the spectroscopy, revealing lines of larger atomic mass number later on
(Pan {\it et al} 2012). The latter effect would arise because of the SN photosphere receding into
deeper layers of the exploding star, thus probing later stages of nuclear burning.

Given the challenges in any search for Pop~III PISNe, it is important to assess the prospects for the
detection of conventional core-collapse explosions from the first stars (Weinmann and Lilly 2005,
Mesinger {\it et al} 2006, Tanaka {\it et al} 2012, Whalen {\it et al} 2013b).
The basic message here is
that, since core-collapse SNe are dimmer than PISNe, one cannot reach the very highest redshifts, maybe
only reaching to $z\lesssim 10$; on the other hand, such events are much more common than PISNe, thus
greatly mitigating the need to conduct very wide observing campaigns. A variation to this theme is
provided by the possibility that metal-free regions persist to much lower redshifts, thus allowing
Pop~III stars, and therefore PISNe, to emerge at $z\lesssim 6$ (Scannapieco {\it et al} 2005).
Such low-redshift Pop~III PISN explosions may be much easier to find, if they exist.

\subsection{High-redshift gamma-ray bursts}

As we have seen, there remain considerable uncertainties with regard to the primordial IMF. It is, however,
likely that at
least a fraction of the first stars collapsed into massive black holes at the end
of their short lives, thus providing viable GRB progenitors.
Traditional sources 
to observe the high-$z$ Universe, such as quasars and Lyman-$\alpha$ emitting galaxies,
suffer from the effects of cosmological dimming, whereas GRB afterglows, if observed
at a fixed time after the trigger, exhibit nearly-flat infrared fluxes out to very high $z$ (Ciardi and Loeb 2000). This counter-intuitive effect arises, because a fixed time interval in the observer frame
translates into an increasingly early time in the source frame. Such earlier times in turn sample
the rapidly decaying GRB lightcurve at the moment of maximal brightness, thus compensating for
the cosmological dimming (increasing luminosity distance). In the hierarchical
setting of cosmic structure formation, earlier times are dominated by lower-mass host systems. The massive hosts
required for quasars and bright galaxies are thus 
increasingly rare at the highest redshifts (Mortlock {\it et al} 2011). 
GRBs, on the other hand, mark the death of individual stars, which can form even in very low-mass systems.
Future missions, such as {\it JANUS}, {\it Lobster}, or {\it SVOM}, promise to fully unleash the potential
of GRB cosmology.

To successfully trigger
a collapsar event, the leading contender for long-duration GRBs (Woosley 1993,
MacFadyen {\it et al} 2001), a number of conditions have to be met. These are
often difficult to fulfill simultaneously 
(Petrovic {\it et al} 2005, Belczynski {\it et al} 2007).
The first requirement, the emergence of BH remnants, is likely fulfilled due to the top-heavy nature
of primordial star formation.
The second key requirement,
that the collapsar progenitor retains enough angular momentum, may be met as well, as is indicated by
the recent work suggesting that
the first stars typically were fast rotators, with
surface rotation speeds of a few 10\% of the break-up value (see section 3.8). 
A third condition is that the relativistic jet, launched by the BH accretion torus,
can escape from the stellar envelope. Though challenging for any progenitor models,
recent work has indicated that jet breakout may be possible in massive Pop~III stars,
even if their extended envelope is not lost prior to the GRB explosion (Komissarov
and Barkov 2010, Suwa and Ioka 2011, Nagakura {\it et al} 2012).
It is, therefore, plausible that all requirements for a collapsar central
engine were in place in the early Universe.

How common were Pop~III GRBs, and do current or planned missions have a fair
chance to detect them?
This question can be addressed within the
following general framework (for details, see Bromm and Loeb 2006):
\begin{equation}
\frac{dN^{\rm obs}_{\rm GRB}}{dz}=\psi^{\rm obs}_{\rm GRB}(z)
\frac{\Delta t_{\rm obs}}{(1+z)}\frac{dV}{dz}\mbox{\ ,}
\end{equation}
where $dN^{\rm obs}_{\rm GRB}$ is the number of GRBs, as observed with
a given instrument, from within a redshift interval $dz$, $\psi^{\rm obs}_{\rm GRB}$ the number of bursts per comoving volume, and the other symbols have
their usual meaning. The connection between the burst number density and
cosmic star formation rate density (SFRD) can be expressed via:
\begin{equation}
\psi^{\rm obs}_{\rm GRB}(z)=\eta_{\rm GRB} \psi_{\ast}(z)
\int_{L_{\rm lim}(z)}^\infty p(L)dL \mbox{\ ,}
\end{equation}
where $\psi_{\ast}(z)$ is the cosmic SFRD, $\eta_{\rm GRB}$ the GRB formation
efficiency, $p(L)$ the GRB luminosity function, and $L_{\rm lim}(z)$ the
minimum intrinsic luminosity required to detect the burst with a given
instrument, from a given redshift.

Most of the intricacies come in when dealing with the efficiency factor.
For simplicity, one could assign a constant value, possibly calibrating
it to the observed Pop~I/II value: $\eta_{\rm GRB}\sim 10^{-9}$ bursts
per unit solar mass (Bromm and Loeb 2006). Within such an idealized model, one
typically estimates that of order 10\% of all {\it Swift} GRBs should
originate from $z>5$, with of order 0.1 Pop~III bursts per year.
Detection of a Pop~III burst may thus lie just outside of the {\it Swift}
capabilities, unless we get lucky.
However, the real situation is likely much more complicated. The GRB efficiency
could well depend on redshift, or on environmental factors, such as the 
metallicity of the host system (Langer and Norman 2006). Since the early modeling
of the GRB redshift distribution, significant refinements have been
added (Daigne {\it et al} 2006, Campisi {\it et al} 2011, deSouza {\it et al} 2011, Ishida {\it et al} 2011, Elliott {\it et al} 2012). Any predictions, though, remain very uncertain, because GRBs are such highly biased tracers
of star formation.
Across a wide range of wavelengths, from the near-IR to radio as well as in the X-ray bands, flux levels are predicted that bring such Pop~III
bursts within reach of existing and planned instruments.
If we can identify these bursts through rapid
follow-up in the near-IR, they will provide us with exquisite background
sources to probe the early IGM (Wang {\it et al} 2012). 

An important unsolved problem in GRB cosmology is how to uniquely identify possible Pop~III bursts.
High-redshift in itself is not sufficient, because different stellar populations will
form contemporaneously, at least at $z\lesssim 15$. Attempts have been made to work out
signatures that rely entirely on the gamma-ray emission, basically derived from the higher
black hole masses expected for Pop~III remnants (M\'{e}sz\'{a}ros and Rees 2010). However,
such diagnostics appear very uncertain, not least because: How would we test or calibrate
such gamma-ray-only markers? The commonly held notion that Pop~III bursts could be
unambiguously identified via the absence of any metal absorption lines in their
afterglow spectra may not work either (see Wang {\it et al} 2012). What are
we then left with? Sightlines toward Pop~III bursts would intersect metal-bubbles produced by neighboring Pop~III stars
that had died somewhat earlier than the GRB progenitor. Such enriched patches would lie at distances exceeding
a few (physical) kpc from the burst. The immediate `near-zone' of the Pop~III GRB, however,
would still be chemically pristine. A unique identifier for Pop~III bursts may thus
be an ensemble of H/He {\it emission} lines, possibly on top of the metal absorption
signal originating farther away from the burst. The emission lines would arise as recombination radiation in the compact H\,{\small II} region powered by the UV-ionizing flux from the GRB afterglow. It is not clear whether the
resulting line fluxes are sufficiently bright
to render them detectable.

\subsection{Nucleosynthesis, Cosmic Archaeology}

The attempt to constrain the properties of the first stars by scrutinizing the chemical abundance
pattern in the atmospheres of extremely old, metal-poor stars in our immediate cosmic neighborhood
has a long and venerable history. What is often termed `stellar archaeology' has resulted in a
rich and growing data set, exhibiting a complex phenomenology, requiring dedicated reviews to 
properly account for it (see Beers and Christlieb 2005, Sneden {\it et al} 2008, Frebel 2013,
Karlsson {\it et al} 2013, and the references therein). We will here only, very briefly, discuss two
issues of particuar promise and importance.

It is often assumed, without being able to firmly prove it, that the most metal-poor stars were
enriched by only one generation of SNe, or even only one prior explosion. The measured abundance
pattern in any given (extreme Pop~II) star can then be matched to a single Pop~III explosion, with
a unique mass of the progenitor (Pop~III) star. Thus, it is, in principle, possible to infer the
Pop~III IMF. This one-to-one mapping is evidently broken in the likely case that more than one
SN contributed to the enrichment. Even then, constraints on the dominant explosion mechanism can
be derived, which in turn yield insight on the primordial IMF. With the advent of large survey
projects, such as the Hamburg-ESO survey (HES) and the Sloan Extension for Galactic Understanding
and Evolution (SEGUE), such a mapping has been attempted, with the tentative result that the majority of
observed metal-poor stars carry the signature of core-collapse SNe (Tumlinson 2006, Heger and Woosley 2010, Joggerst {\it et al} 2010).
Notably, there is no sign for a PISN enrichment, which would manifest itself in a strong elemental odd-even effect,
and the complete absence of any neutron-capture elements. This has often been interpreted as evidence that
the first stars were typically massive, to enable core-collapse events, but probably not very massive ($M_{\ast}
\gtrsim 140 M_{\odot}$), to explain the absence of any PISN signal. However, the apparent lack of
PISN enrichment may be due to an observational selection effect (Karlsson {\it et al} 2008). The idea here is
that even a single PISN can already enrich the neighboring gas to quite high metallicity, of order
$Z\gtrsim 10^{-3} Z_{\odot}$, such that any second-generation stars forming out of this material would
`overshoot' (Greif {\it et al} 2010, Wise {\it et al} 2012). Such stars would then  not be treated as viable candidates for metal-poor stars in the traditional
surveys, which use the lowest metallicities, often the Ca abundance as a proxy, to select for high-resolution spectroscopic follow-up.

\begin{figure}
\includegraphics[width=.7\textwidth]{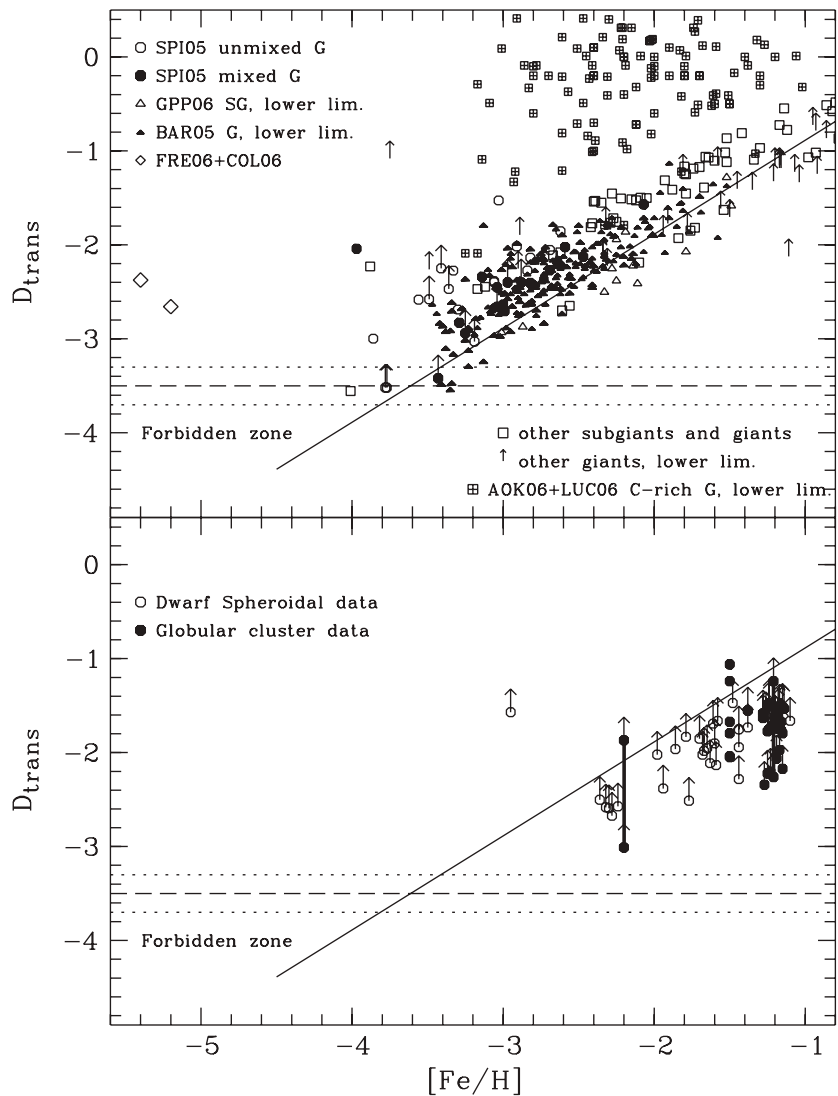}
\caption{Critical metallicity vs. stellar archaeological data.
Shown is the `transition discriminant', $D_{\rm trans}$, which is
a measure of the combined `cooling power' of carbon and oxygen, as
a function of [Fe/H], a logarithmic measure of the iron abundance.
Top panel: Symbols show the data for Milky Way halo giant stars (denoted by G)
and subgiant stars (SG). Bottom panel: Data for stars in Local Group dwarf
spheroidal galaxies and in globular clusters. The minimum, critical $D_{\rm
trans}$ value is marked by the dashed lines, with dotted lines indicating
error bars. The solid lines correspond to the situation simply assuming a
scaled down solar abundance pattern. All stars known pre-2011 fulfill the
fine-structure cooling threshold, but the star recently discovered by 
Caffau {\it et al} (2011) lies in the `Forbidden zone' of the top panel
(not shown here), thus challenging this theory
(adopted from Frebel {\it et al} 2007).
}
\label{fig14}
\end{figure}

A different class of Pop~III SNe are inferred to explain the most iron-poor stars observed yet. Out of 
the 4 stars with [Fe/H]$<-5$, three show strong overabundances in the light CNO elements. Here, we have used
the usual logarithmic notation to describe elemental abundances relative to the solar value, [X/H]$=
\log_{10}(n_{\rm X}/n_{\rm H}) - \log_{10}(n_{\rm X}/n_{\rm H})_{\odot}$, where $n_{\rm X}$ is the number density
of element X, and $n_{\rm H}$ that of hydrogen.
How can one
understand such an enrichment pattern, where large amounts of CNO are produced, but only trace abundances
of all other elements? These peculiar objects are termed carbon-enhanced extremely metal-poor (CEMP) stars,
and they have long been recognized as possibly holding the key to the first stars (Spite {\it et al} 2013).
The CEMP abundance pattern has been explained with the yields from faint, BH-forming SNe (Iwamoto
{\it et al} 2005). Their Pop~III progenitors would have been too massive to trigger conventional Type~II events;
instead, the central region containing the heavier elements would be devoured by the BH, whereas only
the outer envelope, enriched with the lightest, CNO, elements, could escape the deep potential well of the star.
Conceptually, this is a very plausible scenario to explain the CEMP phenomenon. Alternative explanations
invoke AGB enrichment in binary systems (Suda {\it et al} 2004); the problem here is that the known CEMP stars
do not show any sign of binarity.

Recently, a qualitatively new approach has opened up to probe the Pop~III chemical signature:
abundance measurements in high column-density Lyman-$\alpha$ systems at $z\gtrsim 3$.
Among the tantalizing hints are a possible enhancement of carbon
(Cooke {\it et al} 2012), which could be related to the stellar CEMP population, and 
instances of extremely low overall metallicity
(Fumagalli {\it et al} 2011, Simcoe {\it et al} 2012). The latter provide constraints on the
mixing efficiency of Pop~III SN enriched material and the yields of the underlying explosions. 

A second prime use of stellar archaeology is to provide guidance for the theoretical modeling of the
Pop~III -- Pop~II transition, and to test predictions for the critical metallicity (see section~5.2).
More specifically, what is the empirical verdict on the fine-structure line vs. dust-continuum cooling
debate? The fine-structure theory identifies C\,{\small II}, and to a lesser extent O\,{\small I},
as main coolants. This resonates nicely with the prevalence of C-enhanced stars at the lowest
values of [Fe/H]; the resulting [C/H] can then still exceed the predicted critical level. Actually, CEMP stars
are not really `metal-poor'; they are extremely Fe-poor, but otherwise have total metallicities of $Z\gtrsim 10^{-2} 
Z_{\odot}$. Indeed, until
2011, all known metal-poor stars exhibited C and O abundances that are in accordance with the
fine-structure theory (Frebel {\it et al} 2007, see figure~14). This picture has changed in 2011 with the discovery of
the first truly metal-poor star, SDSS~J102915+172927,
with a total metallicity of only $Z\simeq 4.5 \times 10^{-5} Z_{\odot}$ (Caffau {\it et al} 2011).
The formation of SDSS~J102915 is difficult to accommodate with fine-structure line cooling, but can
readily be explained if dust-continuum cooling is invoked (Klessen {\it et al} 2012, Schneider {\it et al} 2012b).
In addition, the observed abundance pattern and inferred dust-to-gas ratio naturally arise in standard
Pop~III core-collapse SN models, even taking into account a reasonable degree of dust destruction in the reverse shock (Schneider {\it et al} 2012).
Still, there remain caveats and open questions. For example, it has been suggested that the extremely low
C abundance observed in SDSS~J102915 is not reflecting the conditions in the birth cloud, but instead is
due to gravitational settling on the MS, and incomplete dredge-up during the subsequent subgiant evolutionary
phase (MacDonald {\it et al} 2013). Another idea has been advanced by Norris {\it et al} (2013), suggesting 
that {\it both} theories for the critical metallicity may play a role, reflected in two clearly distinguishable 
classes of stars at low metallicity: C-enhanced and C-normal. The former class is suggested to arise
from fine-structure cooling, and the latter from dust cooling. Given the complexity of the physics involved,
the developing cross-talk between observation and theory is vital to finally disentangle the role
of metal cooling in the early Universe.

\subsection{Black Hole Remnants}

The possible impact and observational signature of Pop~III BH remnants, both
at high $z$ and in the local Universe, have become a vast subject. To do justice to
it, dedicated reviews are needed, and excellent ones now exist (Greene 2012, Volonteri and
Bellovary 2012, Haiman 2013). We refer the reader to them, and only briefly discuss
some recent developments here.

The cumulative emission from accreting Pop~III remnants at high redshifts could have
contributed to the (soft) cosmic X-ray background (CXB). Intriguingly, a recent analysis
of deep {\it Spitzer}/IRAC data, together with deep {\it Chandra} observations in its
soft bands, originating from the same patch on the sky, has discovered a statistically
significant cross correlation between the unresolved CIB and CXB (Cappelluti {\it et al} 2013).
A Pop~III scenario could naturally explain such a cross-correlation. While the Pop~III
progenitor stars are still alive, their strong ionizing UV radiation would be bottled up in the
still substantially neutral IGM, and thus ultimately reprocessed into Lyman-$\alpha$
photons. These would never be destroyed in the absence of dust, and instead just slowly
diffuse out of the vicinity of the Pop~III star into the general IGM. Redshifting by
about a factor of 10 would then render those photons a part of the CIB. The {\it same}
sources, after having collapsed into a BH upon the Pop~III star's death, would
subsequently contribute X-rays through the accretion of surrounding gas. Again, those
photons may then be redshifted into the soft X-ray bands today, thus contributing part
of the unresolved soft CXB (also see Mesinger {\it et al} 2013, Yue {\it et al} 2013).

A long-standing puzzle concerns the origin and statistics of intermediate-mass BHs, with
masses of a few $100 M_{\odot}$ that would place them between the traditional
regimes of conventional stellar and supermassive BHs (Greene 2012). They are implicated 
as power sources of ultraluminous X-ray sources (ULXs), detected in nearby
galaxies (Fabbiano 2006, Farrell {\it et al} 2009). Given that Pop~III remnants
are predicted to have the right masses for ULX sources, they have been suggested
as candidates (Madau and Rees 2001). However, it is not at all clear how such
remnants could later on capture a binary companion, required to feed the accretion
disk around the BH. However, such Pop~III ULX sources may have existed at high redshifts,
possibly exerting strong feedback effects on the early IGM (Jeon {\it et al} 2012).
The existence and mass distribution of intermediate-mass BHs is crucial to ascertain
whether stellar, possibly Pop~III, remnants are viable candidates to seed the formation
of supermassive BHs at $z\gtrsim 6$, or whether more exotic formation pathways are
required (see the reviews mentioned above for details).

\section{Epilogue}

Understanding the end of the cosmic dark ages and the emergence of the first stars and galaxies
has a special appeal to the imagination, touching some of the fundamental questions
of the ages: What are our cosmic origins? What is our place in the grand sweep of
cosmic evolution? The first star field is entering an exciting period of rapid discovery,
driven by the confluence of supercomputing technology with the advent of
next-generation observational facilities that promise to probe the high-redshift Universe
at the moment of first light. Important challenges remain, however. The theorists need to
push their simulations to ever greater realism, but they also need to formulate robust,
first- or zero-order, empirical tests, providing observers with some clearly defined targets.

When another review on this subject will be written a decade from now, we may well know whether
the key prediction of current theory, that the first stars constituted a distinct stellar
population with a non-standard IMF, is correct. We may, by serendipity, have found
examples of Pop~III explosive deaths in the planned JWST deep-field campaigns, and we may
have made significant headway in sorting out the nucleosynthetic fossil record preserved in the oldest stars
in the Local Group. We may have begun to map the progress of cosmic reionization with
the redshifted 21\,cm radiation from the neutral hydrogen in the early IGM, together with the 
imprint of the first stars in the cosmic infrared background. It is very likely, however, that we
will be surprised in multiple ways, forcing us to revise our theoretical framework in crucial ways.
It will be a privilege to be a part of this ongoing story of discovery.

\section*{Acknowledgments}
VB thanks TAC/UC Berkeley for its hospitality during part of
the work on this review, and Chris McKee for stimulating discussions that
have helped to shape it.
VB is supported by the National Science Foundation grant AST-1009928,
and by NASA through Astrophysics Theory and Fundamental Physics Program 
grant NNX09-AJ33G.

\section*{References}
\smallskip
\begin{harvard}
\item[] Abel T, Anninos P, Zhang Y and Norman M L 1997 {\it New Astron.} {\bf 2} 181
\item[] Abel T, Anninos P, Norman M L and Zhang Y 1998 {\it Astrophys.\ J.} {\bf 508} 518
\item[] Abel T, Bryan G L and Norman M L 2002 {\it Science} {\bf 295} 93
\item[] Alvarez M A, Bromm V and Shapiro P R 2006 {\it Astrophys.\ J.} {\bf 639} 621
\item[] Asplund M, Lambert D L, Nissen P E, Primas F and Smith V V 2006 {\it Astrophys.\ J.} {\bf 644} 229
\item[] Barkana R and Loeb A 2001 {\it Phys.\ Rep.} {\bf 349} 125
\item[] Barkana R and Loeb A 2007 {\it Rep.\ Prog.\ Phys.} {\bf 70} 627
\item[] Beers T C and Christlieb N 2005 {\it Annu.\ Rev.\ Astron.\ Astrophys.} {\bf 43}, 531
\item[] Begelman M C, Rossi E M and Armitage P J 2008 {\it Mon.\ Not.\ R.\ Astron.\ Soc.} {\bf 387} 1649 
\item[] Belczynski K, Bulik T, Heger A and Fryer C 2007 {\it Astrophys.\ J.} {\bf 664} 986
\item[] Bertone G, Hooper D and Silk J 2005 {\it Phys.\ Rep.} {\bf 405} 279
\item[] Biermann L 1950 {\it Z.\ Naturforschung A} {\bf 5} 65
\item[] Birnboim Y and Dekel A 2003 {\it Mon.\ Not.\ R.\ Astron.\ Soc.} {\bf 345} 349
\item[] Bloom J S 2011 {\it What are Gamma-Ray Bursts?} (Princeton: Princeton University Press)
\item[] Bromm V 2000, PhD thesis, Yale Univ.
\item[] Bromm V, Coppi P S and Larson R B 2002 {\it Astrophys.\ J.} {\bf 564} 23
\item[] Bromm V, Ferrara A, Coppi P S and Larson R B 2001 {\it Mon.\ Not.\ R.\ Astron.\ Soc.} {\bf 328} 969
\item[] Bromm V and Larson R B 2004 {\it Annu.\ Rev.\ Astron.\ Astrophys.} {\bf 42} 79
\item[] Bromm V and Loeb A 2002 {\it Astrophys.\ J.} {\bf 575} 111
\item[] Bromm V and Loeb A 2003a {\it Astrophys.\ J.} {\bf 596} 34
\item[] Bromm V and Loeb A 2003b {\it Nature} {\bf 425} 812
\item[] Bromm V and Loeb A 2004 {\it New Astron.} {\bf 9} 353
\item[] Bromm V and Loeb A 2006 {\it Astrophys.\ J.} {\bf 642} 382
\item[] Bromm V and Yoshida N 2011 {\it Annu.\ Rev.\ Astron.\ Astrophys.} {\bf 49} 373
\item[] Bromm V, Yoshida N, Hernquist L and McKee C F 2009 {\it Nature} {\bf 459} 49
\item[] Caffau E {\it et al}  2011 {\it Nature} {\bf 477} 67
\item[] Campisi M A, Maio U, Salvaterra R and Ciardi B 2011 {\it Mon.\ Not.\ R.\ Astron.\ Soc.} {\bf 416} 2760
\item[] Cappelluti N {\it et al} 2013 {\it Astrophys.\ J.} {\bf 769} 68
\item[] Castor J I 2004 {\it Radiation Hydrodynamics} (Cambridge: Cambridge University Press)
\item[] Chatzopoulos E and Wheeler J C 2012 {\it Astrophys.\ J.} {\bf 748} 42
\item[] Chiappini C, Frischknecht U, Meynet G, Hirschi R, Barbuy B, Pignatari M, Decressin T and Maeder A 2011 {\it Nature} {\bf 474} 666
\item[] Ciardi B and Ferrara A 2005 {\it Space Sci.\ Rev.} {\bf 116} 625
\item[] Ciardi B and Loeb A 2000 {\it Astrophys.\ J.} {\bf 540} 687
\item[] Clark P C, Glover S C O and Klessen R S 2008 {\it Astrophys.\ J.} {\bf 672} 757
\item[] Clark P C, Glover S C O, Klessen R S and Bromm V 2011a {\it Astrophys.\ J.} {\bf 727} 110
\item[] Clark P C, Glover S C O, Smith R J, Greif T H, Klessen R S and Bromm V 2011b {\it Science} {\bf 331} 1040
\item[] Cooke R, Pettini M and Murphy M T 2012 {\it Mon.\ Not.\ R.\ Astron.\ Soc.} {\bf 425} 347
\item[] Couchman H M P and Rees M J 1986 {\it Mon.\ Not.\ R.\ Astron.\ Soc.} {\bf 221} 53
\item[] Cucchiara A {\it et al} 2011 {\it Astrophys.\ J.} {\bf 736} 7
\item[] Daigne F, Rossi E M and Mochkovitch R 2011 {\it Mon.\ Not.\ R.\ Astron.\ Soc.} {\bf 372} 1034
\item[] de Souza R S, Yoshida N and Ioka K 2013 {\it Astron.\ Astrophys.} {\bf 533} A32
\item[] de Souza R S, Ciardi B, Maio U and Ferrara A 2013 {\it Mon.\ Not.\ R.\ Astron.\ Soc.} {\bf 428} 2109
\item[] Dessart L, Waldman R, Livne E, Hillier D J and Blondin S 2013 {\it Mon.\ Not.\ R.\ Astron.\ Soc.} {\bf 428} 3227
\item[] Doi K and Susa H 2011 {\it Astrophys.\ J.} {\bf 741} 93
\item[] Dopcke G, Glover S C O, Clark P C and Klessen R S 2013 {\it Astrophys.\ J.} {\bf 766} 103
\item[] Ekstr\"{o}m S, Meynet G, Chiappini C, Hirschi R and Maeder A 2008 {\it Astron.\ Astrophys.} {\bf 489} 685
\item[] Elliott J, Greiner J, Khochfar S, Schady P, Johnson J L and Rau A 2012 {\it Astron.\ Astrophys.} {\bf 539} A113
\item[] Fabbiano G 2006 {\it Annu.\ Rev.\ Astron.\ Astrophys.} {\bf 44}, 323
\item[] Farrell S A, Webb N A, Barret D, Godet O and Rodrigues J M 2009 {\it Nature} {\bf 460} 73
\item[] Federrath C, Sur S, Schleicher D R G, Banerjee R and Klessen R S 2011 {\it Astrophys.\ J.} {\bf 731} 62
\item[] Flower D R, Le~Bourlot J, Pineau des For\^{e}ts G and Roueff E 2000
{\it Mon.\ Not.\ R.\ Astron.\ Soc.} {\bf 314} 753
\item[] Frebel A, Johnson J L and Bromm V 2007 {\it Mon.\ Not.\ R.\ Astron.\ Soc.} {\bf 380} L40
\item[] Frebel A, Johnson J L and Bromm V 2009 {\it Mon.\ Not.\ R.\ Astron.\ Soc.} {\bf 392} L50
\item[] Frebel A 2013, in
{\it The First Galaxies: Theoretical Predictions and Observational Clues},
ed. T Wiklind, B Mobasher and V Bromm (Berlin: Springer), 377
\item[] Freeman K and Bland-Hawthorn J 2002 {\it Annu.\ Rev.\ Astron.\ Astrophys.} {\bf 40}, 487
\item[] Freese K, Bodenheimer P, Spolyar D and Gondolo P 2008 {\it Astrophys.\ J.} {\bf 685} L101
\item[] Freese K, Ilie C, Spolyar D, Valluri M and Bodenheimer P 2010 {\it Astrophys.\ J.} {\bf 716} 1397
\item[] Frommhold, L 1994 {\it Collision-induced Absorption in Gases} (Cambridge: Cambridge University Press)
\item[] Fuller T M and Couchman H M P 2000 {\it Astrophys.\ J.} {\bf 544} 6
\item[] Fumagalli M, O'Meara J M and Prochaska J X 2011 {\it Science} {\bf 334} 1245
\item[] Furlanetto S R, Oh S P and Briggs F H 2006 {\it Phys.\ Rep.} {\bf 433} 181
\item[] Gall C, Hjorth J and Andersen A C 2011 {\it Astron.\ Astrophys.\ Rev.} {\bf 19} 43
\item[] Galli D and Palla F 1998 {\it Astron.\ Astrophys.} {\bf 335} 403
\item[] Galli D and Palla F 2002 {\it Planet.\ Space Sci.} {\bf 50} 1197
\item[] Galli D and Palla F 2013 {\it Annu.\ Rev.\ Astron.\ Astrophys.} {\bf 51} 163 
\item[] Gammie C F 2001 {\it Astrophys.\ J.} {\bf 553} 174
\item[] Gao L and Theuns T 2007 {\it Science} {\bf 317} 1527
\item[] Gao L, Yoshida N, Abel T, Frenk C S, Jenkins A and Springel V 2007 {\it Mon.\ Not.\ R.\ Astron.\ Soc.} {\bf 378} 449
\item[] Gardner J P 2006 {\it Space Sci.\ Rev.} {\bf 123} 485
\item[] Glover S C O 2005 {\it Space Sci.\ Rev.} {\bf 117} 445
\item[] Glover S C O 2013, in
{\it The First Galaxies: Theoretical Predictions and Observational Clues},
ed. T Wiklind, B Mobasher and V Bromm (Berlin: Springer), 103
\item[] Glover S C O and Abel T 2008 {\it Mon.\ Not.\ R.\ Astron.\ Soc.} {\bf 388} 1627
\item[] Greene J E 2012 {\it Nat.\ Commun.} {\bf 3} 1304
\item[] Greif T H, Johnson J L, Klessen R S and Bromm V 2008 {\it Mon.\ Not.\ R.\ Astron.\ Soc.} {\bf 387} 1021
\item[] Greif T H, Glover S C O, Bromm V and Klessen R S 2010 {\it Astrophys.\ J.} {\bf 716} 510
\item[] Greif T H, Springel V, White S D M, Glover S C O, Clark P C, Smith R J, Klessen R S and Bromm V 2011a {\it Astrophys.\ J.} {\bf 737} 75
\item[] Greif T H, White S D M, Klessen R S and Springel V 2011b {\it Astrophys.\ J.} {\bf 736} 147
\item[] Greif T H, Bromm V, Clark P C, Glover S C O, Smith R J, Klessen R S, Yoshida N and Springel V 2012 {\it Mon.\ Not.\ R.\ Astron.\ Soc.} {\bf 424} 399
\item[] Greif T H, Springel V and Bromm V 2013 {\it Mon.\ Not.\ R.\ Astron.\ Soc.} {\bf 434} 3408
\item[] Haiman Z, Thoul A A and Loeb A 1996 {\it Astrophys.\ J.} {\bf 464} 523
\item[] Haiman Z, Rees M J and Loeb A 1997 {\it Astrophys.\ J.} {\bf 476} 458
\item[] Haiman Z 2013, in
{\it The First Galaxies: Theoretical Predictions and Observational Clues},
ed. T Wiklind, B Mobasher and V Bromm (Berlin: Springer), 293
\item[] Heger A and Woosley S E 2010 {\it Astrophys.\ J.} {\bf 724} 341
\item[] Hirano S, Umeda H and Yoshida N 2011 {\it Astrophys.\ J.} {\bf 736} 58
\item[] Hirano S and Yoshida N 2013 {\it Astrophys.\ J.} {\bf 763} 52
\item[] Hollenbach D and  Salpeter E E 1971 {\it Astrophys.\ J.} {\bf 163} 155
\item[] Hosokawa T, Omukai K, Yoshida N and Yorke H W 2011 {\it Science} {\bf 334} 1250
\item[] Hosokawa T, Yoshida N, Omukai K and Yorke H W 2012 {\it Astrophys.\ J.} {\bf 760} L37
\item[] Hummel J A, Pawlik A H, Milosavljevi\'{c} M and Bromm V 2012 {\it Astrophys.\ J.} {\bf 755} 72
\item[] Iben I 1983 {\it Mem.\ Soc.\ Astron.\ Ital.} {\bf 54} 321
\item[] Ilie C, Freese K, Valluri M, Iliev I T and Shapiro P R 2012 {\it Mon.\ Not.\ R.\ Astron.\ Soc.} {\bf 422} 2164
\item[] Inayoshi K and Omukai K 2011 {\it Mon.\ Not.\ R.\ Astron.\ Soc.} {\bf 416} 2748
\item[] Iocco F 2008 {\it Astrophys.\ J.} {\bf 677} L1
\item[] Iocco F, Bressan A, Ripamonti E, Schneider R, Ferrara A and Marigo P 2008 {\it Mon.\ Not.\ R.\ Astron.\ Soc.} {\bf 390} 1655
\item[] Ishida E E O, de Souza R S and Ferrara A 2011 {\it Mon.\ Not.\ R.\ Astron.\ Soc.} {\bf 418} 500
\item[] Iwamoto N, Umeda H, Tominaga N, Nomoto K and Maeda K 2005 {\it Science} {\bf 309} 451
\item[] Jappsen A-K, Mac Low M-M, Glover S C O, Klessen R S and Kitsionas S 2009a {\it Astrophys.\ J.} {\bf 694} 1161
\item[] Jappsen A-K, Klessen R S, Glover S C O and Mac Low M-M 2009b {\it Astrophys.\ J.} {\bf 696} 1065
\item[] Jasche J, Ciardi B and En{\ss}lin T A 2007 {\it Mon.\ Not.\ R.\ Astron.\ Soc.} {\bf 380} 417
\item[] Jeon M, Pawlik A H, Greif H, Glover S C O, Bromm V, Milosavljevi\'{c} M and Klessen R S 2012 {\it Astrophys.\ J.} {\bf 754} 34
\item[] Joggerst C C, Almgren A, Bell J, Heger A, Whalen D and Woosley S E 2010 {\it Astrophys.\ J.} {\bf 709} 11
\item[] Johnson J L and Bromm V 2006 {\it Mon.\ Not.\ R.\ Astron.\ Soc.} {\bf 366} 247
\item[] Johnson J L, Greif T H and Bromm V 2008 {\it Mon.\ Not.\ R.\ Astron.\ Soc.} {\bf 388} 26
\item[] Johnson J L and Khochfar S 2011 {\it Mon.\ Not.\ R.\ Astron.\ Soc.} {\bf 413} 1184
\item[] Johnson J L, Whalen D J, Even W, Fryer C L, Heger A, Smidt J and Chen K-J 2013 {\it Astrophys.\ J.}
in press ({\it Preprint} arXiv:1304.4601)
\item[] Kaplinghat M, Chu M, Haiman Z, Holder G P, Know L and Skordis C 2003 {\it Astrophys.\ J.} {\bf 583} 24
\item[] Kasen D, Woosley S E and Heger A 2011 {\it Astrophys.\ J.} {\bf 734} 102
\item[] Karlsson T, Johnson J L and Bromm V 2008 {\it Astrophys.\ J.} {\bf 679} 6
\item[] Karlsson T, Bromm V and Bland-Hawthorn J 2013 {\it Rev.\ Mod.\ Phys.} {\bf 85} 809
\item[] Kashlinsky A 2005 {\it Phys.\ Rep.} {\bf 409} 361
\item[] Kere\v{s} D, Katz N, Weinberg D H and Dav\'{e} R 2005 {\it Mon.\ Not.\ R.\ Astron.\ Soc.} {\bf 363} 2
\item[] Kitayama T, Yoshida N, Susa H and Umemura M 2004 {\it Astrophys.\ J.} {\bf 613} 631
\item[] Klessen R S, Glover S C O and Clark P C 2012 {\it Mon.\ Not.\ R.\ Astron.\ Soc.} {\bf 421} 3217
\item[] Komatsu E {\it et al} 2011 {\it Astrophys.\ J.\ Suppl.} {\bf 192} 18
\item[] Komissarov S S and Barkov M V 2010 {\it Mon.\ Not.\ R.\ Astron.\ Soc.} {\bf 402} L25
\item[] Kratter K M, Matzner C D, Krumholz M R and Klein R I 2010 {\it Astrophys.\ J.} {\bf 708} 1585
\item[] Kreckel H, Bruhns H, \v{C}\'{i}\v{z}ek M, Glover S C O, Miller K A, Urbain X and Savin D W 2010 {\it Science} {\bf 329} 69
\item[] Kritsuk A G, Norman M L and Wagner R 2011, {\it Astrophys.\ J.} {\bf 727}, L20
\item[] Krumholz M R, Klein R I and McKee C F 2007 {\it Astrophys.\ J.} {\bf 656} 959
\item[] Krumholz M R, Klein R I, McKee C F, Offner, S S R and Cunningham A J 2009 {\it Science} {\bf 323} 754
\item[] Kudritzki R P 2002 {\it Astrophys.\ J.} {\bf 577} 389
\item[] Kulsrud R M 2005 {\it Plasma Physics for Astrophysics} (Princeton: Princeton University Press)
\item[] Lamb D Q and Reichart D E 2000 {\it Astrophys.\ J.} {\bf 536} 1
\item[] Langer N 2012 {\it Annu.\ Rev.\ Astron.\ Astrophys.} {\bf 50} 107
\item[] Langer N and Norman C A 2006 {\it Astrophys.\ J.} {\bf 638} L63
\item[] Larson R B 1969 {\it Mon.\ Not.\ R.\ Astron.\ Soc.} {\bf 145} 271
\item[] Larson R B 1981 {\it Mon.\ Not.\ R.\ Astron.\ Soc.} {\bf 194} 809
\item[] Larson R B 1998 {\it Mon.\ Not.\ R.\ Astron.\ Soc.} {\bf 301} 569
\item[] Larson R B 2003 {\it Rep.\ Prog.\ Phys.} {\bf 66} 1651
\item[] Latif M A, Schleicher D R G, Schmidt W and Niemeyer J 2013a {\it Mon.\ Not.\ R.\ Astron.\ Soc.} {\bf 432} 668
\item[] Latif M A, Schleicher D R G, Schmidt W and Niemeyer J 2013b {\it Mon.\ Not.\ R.\ Astron.\ Soc.} {\bf 433} 1607
\item[] Loeb A 2010 {\it How did the First Stars and Galaxies From?} (Princeton: Princeton University Press)
\item[] Loeb A and Furlanetto S R 2013 {\it The First Galaxies in the Universe} (Princeton: Princeton University Press)
\item[] Low C and Lynden-Bell D 1976 {\it Mon.\ Not.\ R.\ Astron.\ Soc.} {\bf 176} 367
\item[] MacDonald J, Lawlor T M, Anilmis N and Rufo N F 2013 {\it Mon.\ Not.\ R.\ Astron.\ Soc.} {\bf 431} 1425
\item[] MacFadyen A I and Woosley S E 1999 {\it Astrophys.\ J.} {\bf 524} 262
\item[] MacFadyen A I, Woosley S E and Heger A 2001 {\it Astrophys.\ J.} {\bf 550} 410
\item[] Machacek M E, Bryan G L and Abel T 2001 {\it Astrophys.\ J.} {\bf 548} 509
\item[] Machida M N, Omukai K, Matsumoto T and Inutsuka S 2006 {\it Astrophys.\ J.} {\bf 647} L1
\item[] Machida M N, Omukai K, Matsumoto T and Inutsuka S 2008a {\it Astrophys.\ J.} {\bf 677} 813
\item[] Machida M N, Matsumoto T and Inutsuka S 2008b {\it Astrophys.\ J.} {\bf 685} 690
\item[] Mackey J, Bromm V and Hernquist L 2003 {\it Astrophys.\ J.} {\bf 586} 1
\item[] Mac Low M-M and Klessen R S 2004 {\it Rev.\ Mod.\ Phys.} {\bf 76} 125
\item[] Madau P and Rees M J 2001 {\it Astrophys.\ J.} {\bf 551} L27
\item[] Maeder A and Meynet G 2012 {\it Rev.\ Mod.\ Phys.} {\bf 84} 25
\item[] Maio U, Koopmans L V E and Ciardi B 2011 {\it Mon.\ Not.\ R.\ Astron.\ Soc.} {\bf 412} L40
\item[] McGreer I D and Bryan G L 2008 {\it Astrophys.\ J.} {\bf 685} 8
\item[] McKee C F and Ostriker E C 2007 {\it Annu.\ Rev.\ Astron.\ Astrophys.} {\bf 45} 565
\item[] McKee C F and Tan J C 2008 {\it Astrophys.\ J.} {\bf 681} 771
\item[] Meiksin A A 2009 {\it Rev.\ Mod.\ Phys.} {\bf 81} 1405
\item[] Mesinger A, Johnson B D and Haiman Z 2006 {\it Astrophys.\ J.} {\bf 637} 80
\item[] Mesinger A, Ferrara A and Spiegel D S 2013 {\it Mon.\ Not.\ R.\ Astron.\ Soc.} {\bf 431} 621
\item[] M\'{e}sz\'{a}ros P and Rees M J 2010 {\it Astrophys.\ J.} {\bf 715} 967
\item[] Miralda-Escud\'{e} J 2003 {\it Science} {\bf 300} 1904
\item[] Mortlock D J {\it et al} 2011 {\it Nature} {\bf 474} 616
\item[] Nagakura H, Suwa Y and Ioka K 2012 {\it Astrophys.\ J.} {\bf 754} 85
\item[] Naoz S, Noter S and Barkana R 2006 {\it Mon.\ Not.\ R.\ Astron.\ Soc.} {\bf 373} L98
\item[] Naoz S, Yoshida N and Gnedin N Y 2012 {\it Astrophys.\ J.} {\bf 747} 128
\item[] Naoz S, Yoshida N and Gnedin N Y 2013 {\it Astrophys.\ J.} {\bf 763} 27
\item[] Natarajan A, Tan J C and O'Shea B W 2009 {\it Astrophys.\ J.} {\bf 692} 574
\item[] Norris J E {\it et al}  2013 {\it Astrophys.\ J.} {\bf 762} 28
\item[] Omukai K 2000 {\it Astrophys.\ J.} {\bf 534} 809
\item[] Omukai K and Inutsuka S 2002 {\it Mon.\ Not.\ R.\ Astron.\ Soc.} {\bf 332} 59
\item[] Omukai K and Nishi R 1998 {\it Astrophys.\ J.} {\bf 508} 141
\item[] Omukai K and Palla F 2001 {\it Astrophys.\ J.} {\bf 561} L55
\item[] Omukai K and Palla F 2003 {\it Astrophys.\ J.} {\bf 589} 677
\item[] O'Shea B W and Norman M L 2007 {\it Astrophys.\ J.} {\bf 654} 66
\item[] O'Shea B W, McKee C F, Heger A and Abel T 2008, in AIP Conf. Proc. 990,
{\it First Stars III}, ed. B W O'Shea and A Heger,
Melville, NY: AIP, 13
\item[] Palla F, Salpeter E E and Stahler S W 1983 {\it Astrophys.\ J.} {\bf 271} 632
\item[] Pan T, Kasen D and Loeb A 2012 {\it Mon.\ Not.\ R.\ Astron.\ Soc.} {\bf 422} 2701
\item[] Pawlik A H, Milosavljevi\'{c} M and Bromm V 2011 {\it Astrophys.\ J.} {\bf 731} 54
\item[] Penston M V 1969 {\it Mon.\ Not.\ R.\ Astron.\ Soc.} {\bf 144} 425
\item[] Petrovic J, Langer N, Yoon S-C and Heger A 2005 {\it Astron.\ Astrophys.} {\bf 435} 247
\item[] Planck Collaboration 2013 {\it Astron.\ Astrophys.} submitted ({\it Preprint} arXiv:1303.5076)
\item[] Prieto J, Padoan P, Jimenez R and Infante L 2011 {\it Astrophys.\ J.} {\bf 731} L38
\item[] Pudritz R E 1981 {\it Mon.\ Not.\ R.\ Astron.\ Soc.} {\bf 195} 897
\item[] Rees M J 1976 {\it Mon.\ Not.\ R.\ Astron.\ Soc.} {\bf 176} 483
\item[] Rees M J and Ostriker J P 1977 {\it Mon.\ Not.\ R.\ Astron.\ Soc.} {\bf 179} 541
\item[] Rees M J 2000 {\it New Perspectives in Astrophysical Cosmology} (Cambridge: Cambridge University Press)
\item[] Ripamonti E and Abel T 2004 {\it Mon.\ Not.\ R.\ astron.\ Soc.} {\bf 348} 1019
\item[] Ripamonti E, Mapelli M and Ferrara A 2007 {\it Mon.\ Not.\ R.\ astron.\ Soc.} {\bf 375} 1399
\item[] Ripamonti E, Iocco F, Ferrara A, Schneider R, Bressan A and Marigo P 2010 {\it Mon.\ Not.\ R.\ astron.\ Soc.} {\bf 406} 2605
\item[] Robertson B E, Ellis R S, Dunlop J S, McLure R J and Stark D P 2010 {\it Nature} {\bf 468} 49
\item[] Rollinde E, Vangioni E and Olive K A 2005 {\it Astrophys.\ J.} {\bf 627} 666
\item[] Rollinde E, Vangioni E and Olive K A 2006 {\it Astrophys.\ J.} {\bf 651} 658
\item[] Rydberg C-E, Zackrisson E, Lundqvist P and Scott P 2013 {\it Mon.\ Not.\ R.\ astron.\ Soc.} {\bf 429} 3658
\item[] Safranek-Shrader C, Bromm V and Milosavljevi\'{c} M 2010 {\it Astrophys.\ J.} {\bf 723} 1568
\item[] Safranek-Shrader C, Agarwal M, Federrath C, Dubey A, Milosavljevi\'{c} and Bromm V 2012 {\it Mon.\ Not.\ R.\ Astron.\ Soc.} {\bf 426} 1159
\item[] Salpeter E E 1955 {\it Astrophys.\ J.} {\bf 121} 161
\item[] Salvaterra R {\it et al} 2009 {\it Nature} {\bf 461} 1258
\item[] Santoro F and Shull J M 2006 {\it Astrophys.\ J.} {\bf 643} 26
\item[] Saslaw W C and Zipoy D 1967 {\it Nature} {\bf 216} 976
\item[] Scannapieco E, Madau P, Woosley S E, Heger A and Ferrara A 2005 {\it Astrophys.\ J.} {\bf 633} 1031
\item[] Schleicher D R G, Banerjee R, Sur S, Arshakian T G, Klessen R S, Beck R and Spaans M 2010 {\it Astron.\ Astrophys.} {\bf 522} 115
\item[] Schlickeiser R 2002 {\it Cosmic Ray Astrophysics} (Berlin: Springer)
\item[] Schneider R, Ferrara A, Natarajan P and Omukai K 2002 {\it Astrophys.\ J.} {\bf 571} 30
\item[] Schneider R and Omukai K 2010 {\it Mon.\ Not.\ R.\ Astron.\ Soc.} {\bf 402} 429
\item[] Schneider R, Omukai K, Inoue A K and Ferrara A 2006 {\it Mon.\ Not.\ R.\ Astron.\ Soc.} {\bf 369} 1437
\item[] Schneider R, Omukai K, Bianchi S and Valiante R 2012a {\it Mon.\ Not.\ R.\ Astron.\ Soc.} {\bf 419} 1566
\item[] Schneider R, Omukai K, Limongi M, Ferrara A, Salvaterra R, Chieffi A and Bianchi S 2012b {\it Mon.\ Not.\ R.\ Astron.\ Soc.} {\bf 423} L60
\item[] Schober J, Schleicher D, Federrath C, Glover S, Klessen R S and Banerjee R 2012 {\it Astrophys.\ J.} {\bf 754} 99
\item[] Shakura N I and Sunyaev R A 1973 {\it Astron.\ Astrophys.} {\bf 24} 337
\item[] Shchekinov Y A and Vasiliev E O 2004 {\it Astron.\ Astrophys.} {\bf 419} 19
\item[] Shu F H 1977 {\it Astrophys.\ J.} {\bf 214} 488
\item[] Silk J 1977 {\it Astrophys.\ J.} {\bf 211} 638
\item[] Silk J and Langer M 2006 {\it Mon.\ Not.\ R.\ Astron.\ Soc.} {\bf 371} 444
\item[] Simcoe R A, Sullivan P W, Cooksey K L, Kao M M, Matejek M S and Burgasser A J 2012 {\it Nature} {\bf 492} 79
\item[] Sivertsson S and Gondolo P 2011 {\it Astrophys.\ J.} {\bf 729} 51
\item[] Smith B D, Turk M J, Sigurdsson S, O'Shea B W and Norman M L 2009 {\it Astrophys.\ J.} {\bf 691} 441
\item[] Smith R J, Glover S C O, Clark P C, Greif T H and Klessen R S 2011 {\it Mon. Not. R. Astron. Soc.} {\bf 414} 3633
\item[] Smith R J, Hosokawa T, Omukai K, Glover S C O and Klessen R S 2012a {\it Mon. Not. R. Astron. Soc.} {\bf 424} 457
\item[] Smith R J, Iocco F, Glover S C O, Scleicher D R G, Klessen R S, Hirano S and Yoshida N 2012b {\it Astrophys.\ J.} {\bf 761} 154
\item[] Sneden C, Cowan J J and Gallino R 2008 {\it Annu.\ Rev.\ Astron.\ Astrophys.} {\bf 46} 241
\item[] Spite F and Spite M 1982 {\it Astron.\ Astrophys.} {\bf 115} 357
\item[] Spite M, Caffau E, Bonifacio P, Spite F, Ludwig H-G, Plez B and Christlieb N 2013 {\it Astron.\ Astrophys.} {\bf 552} A107
\item[] Spolyar D, Freese K and Gondolo P 2008 {\it Phys.\ Rev.\ Lett.} {\bf 100} 051101
\item[] Springel V 2010 {\it Mon.\ Not.\ R.\ Astron.\ Soc.} {\bf 401} 791
\item[] Stacy A and Bromm V 2007 {\it Mon.\ Not.\ R.\ Astron.\ Soc.} {\bf 382} 229
\item[] Stacy A, Greif T H and Bromm V 2010 {\it Mon.\ Not.\ R.\ Astron.\ Soc.} {\bf 403} 45
\item[] Stacy A, Bromm V and Loeb A 2011a {\it Astrophys.\ J.} {\bf 730} L1
\item[] Stacy A, Bromm V and Loeb A 2011b {\it Mon.\ Not.\ R.\ Astron.\ Soc.} {\bf 413} 543
\item[] Stacy A, Pawlik A H, Bromm V and Loeb A 2012a {\it Mon.\ Not.\ R.\ Astron.\ Soc.} {\bf 421} 894
\item[] Stacy A, Greif T H and Bromm V 2012b {\it Mon.\ Not.\ R.\ Astron.\ Soc.} {\bf 422} 290
\item[] Stacy A, Greif T H, Klessen R S, Bromm V and Loeb A 2013a {\it Mon.\ Not.\ R.\ Astron.\ Soc.} {\bf 431} 1470
\item[] Stacy A and Bromm V 2013b {\it Mon.\ Not.\ R.\ Astron.\ Soc.} {\bf 433} 1094
\item[] Stahler S W, Palla F and Salpeter E E 1986 {\it Astrophys.\ J.} {\bf 302} 590
\item[] Stahler S W and Palla F 2004 {\it The Formation of Stars} (Weinheim: Wiley-VCH)
\item[] Stiavelli M 2009 {\it From First Light to Reionization: The End of the Dark Ages} (Weinheim: Wiley-VCH)
\item[] Suda T, Aikawa M, Machida M N, Fujimoto M Y and Iben I 2004 {\it Astrophys.\ J.} {\bf 611} 476
\item[] Suda T {\it et al} 2013 {\it Mon.\ Not.\ R.\ Astron.\ Soc.} {\bf 432} L46
\item[] Sur S, Schleicher D R G, Banerjee R, Federrath C and Klessen R S 2010 {\it Astrophys.\ J.} {\bf 721} L134
\item[] Suwa Y and Ioka K 2011 {\it Astrophys.\ J.} {\bf 726} 107
\item[] Tan J C and Blackman E G 2004 {\it Astrophys.\ J.} {\bf 603} 401
\item[] Tan J C and McKee C F 2004 {\it Astrophys.\ J.} {\bf 603} 383
\item[] Tanaka M, Moriya T J, Yoshida N and Nomoto K 2012 {\it Mon.\ Not.\ R.\ Astron.\ Soc.} {\bf 422} 2675
\item[] Tanvir N R {\it et al} 2009 {\it Nature} {\bf 461} 1254
\item[] Tegmark M, Silk J, Rees M J, Blanchard A, Abel T and Palla F 1997 {\it Astrophys.\ J.} {\bf 474} 1
\item[] Tielens A G G M 2005 {\it The Physics and Chemistry of the Interstellar Medium} (Cambridge: Cambridge University Press)
\item[] Truelove J K, Klein R I, McKee C F, Holliman J H, Howell L H, Greenough J A and Woods D T 1998 {\it Astrophys.\ J.} {\bf 495} 821
\item[] Tseliakhovich D and Hirata C M 2010 {\it Phys.\ Rev.}\ D {\bf 82} 083520
\item[] Tseliakhovich D, Barkana R and Hirata C M 2011 {\it Mon.\ Not.\ R.\ Astron.\ Soc.} {\bf 418} 906
\item[] Tsuribe T and Omukai K 2006 {\it Astrophys.\ J.} {\bf 642} L61
\item[] Tumlinson J 2006 {\it Astrophys.\ J.} {\bf 641} 1
\item[] Turk M J, Abel T and O'Shea B 2009 {\it Science} {\bf 325} 601
\item[] Turk M J, Clark P C, Glover S C O, Greif T H, Abel T, Klessen R S and Bromm V 2011 {\it Astrophys.\ J.} {\bf 726} 55
\item[] Turk M J, Oishi J S, Abel T and Bryan G L 2012 {\it Astrophys.\ J.} {\bf 745} 154
\item[] Uehara H, Susa H, Nishi R, Yamada M and Nakamura T 1996 {\it Astrophys.\ J.} {\bf 473} L95
\item[] Visbal E, Barkana R, Fialkov A, Tseliakhovich D and Hirata C M 2012 {\it Nature} {\bf 487} 70
\item[] Volonteri M and Bellovary J 2012 {\it Rep.\ Prog.\ Phys.} {\bf 75} 124901
\item[] Wang F Y, Bromm V, Greif T H, Stacy A, Dai Z G, Loeb A and Cheng K S 2012 {\it Astrophys.\ J.} {\bf 760} 27
\item[] Weinmann S M and Lilly S J 2005 {\it Astrophys.\ J.} {\bf 624} 526
\item[] Whalen D, Abel T and Norman M L 2004 {\it Astrophys.\ J.} {\bf 610} 14
\item[] Whalen D, Bromm V and Yoshida N, eds. 2010
{\it The First Stars and Galaxies: Challenges for the
Next Decade.} AIP Conf. Proc. 1294. Melville, NY: AIP
\item[] Whalen D J, Fryer C L, Holz D E, Heger A, Woosley S E, Stiavelli M, Even W and Frey L H 2013a {\it Astrophys.\ J.} {\bf 762} L6
\item[] Whalen D J, Joggerst C C, Fryer C L, Stiavelli M, Heger A and Holz D E 2013b {\it Astrophys.\ J.} {\bf 768} 95
\item[] Whitehouse S C and Bate M R 2006 {\it Mon.\ Not.\ R.\ Astron.\ Soc.} {\bf 367} 32
\item[] Widrow L M, Ryu D, Schleicher D R G, Subramanian K, Tsagas C G and Treumann R A 2012 {\it Space Sci.\ Rev.} {\bf 166} 37
\item[] Wiklind T, Mobasher B and Bromm V, eds. 2013
{\it The First Galaxies: Theoretical Predictions and Observational
Clues} (Berlin: Springer)
\item[] Wise J H and Abel T 2007 {\it Astrophys.\ J.} {\bf 665} 899
\item[] Wise J H and Abel T 2008 {\it Astrophys.\ J.} {\bf 685} 40
\item[] Wise J H, Turk M J, Norman M L and Abel T 2012 {\it Astrophys.\ J.} {\bf 745} 50
\item[] Wolcott-Green J and Haiman Z 2011 {\it Mon.\ Not.\ R.\ Astron.\ Soc.} {\bf 412} 2603
\item[] Wolcott-Green J, Haiman Z and Bryan G L 2011 {\it Mon.\ Not.\ R.\ Astron.\ Soc.} {\bf 418} 838
\item[] Woosley S E 1993 {\it Astrophys.\ J.} {\bf 405} 273
\item[] Woosley S E and Heger A 2006 {\it Astrophys.\ J.} {\bf 637} 914
\item[] Xu H, O'Shea B W, Collins D C, Norman M L, Li H and Li S 2008 {\it Astrophys.\ J.} {\bf 688} L57
\item[] Yoon S-C and Langer N 2005 {\it Astron.\ Astrophys.} {\bf 443} 643
\item[] Yoon S-C, Iocco F and Akiyama S 2008 {\it Astrophys.\ J.} {\bf 688} L1
\item[] Yoon S-C, Dierks A and Langer N 2012 {\it Astron.\ Astrophys.} {\bf 542} A113
\item[] Yoshida N, Abel T, Hernquist L and Sugiyama N 2003a {\it Astrophys.\ J.} {\bf 592} 645
\item[] Yoshida N, Sokasian A, Hernquist L and Springel V 2003b {\it Astrophys.\ J.} {\bf 591} L1
\item[] Yoshida N, Omukai K, Hernquist L and Abel T 2006 {\it Astrophys.\ J.} {\bf 652} 6
\item[] Yoshida N, Oh S P, Kitayama T and Hernquist L 2007 {\it Astrophys.\ J.} {\bf 663} 687
\item[] Yoshida N, Omukai K and Hernquist L 2008 {\it Science} {\bf 321} 669
\item[] Yoshii Y 1981 {\it Astron.\ Astrophys.} {\bf 97} 280
\item[] Yue B, Ferrara A, Salvaterra R, Xu Y and Chen X 2013 {\it Mon.\ Not.\ R.\ Astron.\ Soc.} {\bf 433} 1556
\item[] Zackrisson E {\it et al} 2010 {\it Mon.\ Not.\ R.\ Astron.\ Soc.} {\bf 407} L74
\item[] Zinnecker H and Yorke H W 2007 {\it Annu.\ Rev.\ Astron.\ Astrophys.} {\bf 45} 481
\end{harvard}

\end{document}